\documentclass[fleqn,usenatbib]{mnras}

\usepackage{newtxtext,newtxmath}
\usepackage[T1]{fontenc}

\DeclareRobustCommand{\VAN}[3]{#2}
\let\VANthebibliography\thebibliography
\def\thebibliography{\DeclareRobustCommand{\VAN}[3]{##3}\VANthebibliography}

\usepackage{graphicx}	% Including figure files
\usepackage{amsmath}	% Advanced maths commands
\usepackage{orcidlink}  % To create a linked ORCiD logo for authors
\usepackage{multirow}   % To allow multi-row items in tables
\usepackage{textgreek}  % To include non-italic greek letters
\usepackage[export]{adjustbox} %To get a right aligned image

\usepackage{color,soul} % allows \hl{} to highlight text
\usepackage{threeparttable}

\defcitealias{MH2008}{MH08}

\newcommand{\ts}{\textsuperscript}

\title[HD\,28471 planetary system]{HD\,28471: a near-resonant compact multiplanet system with a possible cold giant planet}
\author[A. T. Stevenson et al.]{
A. T. Stevenson$^{1,2}$\thanks{E-mail: a.t.stevenson@bham.ac.uk}\orcidlink{0000-0003-2399-7619}, % can i stay a visiting researcher to keep OU email since im collaborating? I think it is lots of admin...
C. A. Haswell$^{1}$\orcidlink{0000-0002-8050-1897}, 
J. R. Barnes$^{1}$\orcidlink{0000-0001-6105-2902}, 
M. R. Standing$^{3}$\orcidlink{0000-0002-7608-8905}, 
J. K. Barstow$^{1}$\orcidlink{0000-0003-3726-5419},\newauthor
~~Z. O. B. Ross$^{1}$\orcidlink{0000-0002-3213-9643},
A. V. Freckelton$^{2}$\orcidlink{0009-0007-1053-0004},
D. Staab$^{1}$ \\
\\
$^{1}$School of Physical Sciences, The Open University, Milton Keynes MK7 6AA, UK \\
$^{2}$School of Physics \& Astronomy, University of Birmingham, Edgbaston, Birmingham B15 2TT, UK\\
$^{3}$European Space Agency (ESA), European Space Astronomy Centre (ESAC), Camino Bajo del Castillo s/n, E-28692 Villanueva de la Ca\~nada, Madrid, Spain
}

\date{Accepted 2025 August 20. Received 2025 August 08; in original form 2025 March 05}

\pubyear{2025}

\begin{document}
\label{firstpage}
\pagerange{\pageref{firstpage}--\pageref{lastpage}}
\maketitle

% Abstract of the paper
\begin{abstract}
%250 limit for papers, 200 limit for letters. No references.
We present radial velocity measurements of the star HD\,28471, observed by HARPS at the ESO 3.6\,m telescope over a baseline of $\sim19$\,years. We have searched for planetary companions to HD\,28471 using \textsc{kima}, a trans-dimensional diffusive nested sampling algorithm where the number of planetary signals is explored as a free parameter. We detect a compact system of three planets, with signals in the preferred solution corresponding to orbits of $P\sim3.16,~6.12,~\textrm{and~}11.68$~d. These planets lie firmly in the super-Earth and sub-Neptune mass regime, with (minimum) masses of $3.7,~5.7,~\textrm{and~}4.9$~M$_{\earth}$, respectively. A long-period ($\sim1500$~d) signal is also strongly detected. Assessment of activity indicator periodicities and RV correlations suggests that the three short-period signals are genuine planets, but casts doubt upon the nature of the long-period signal. The origin may be a short stellar magnetic cycle, though additional data are required to fully sample the periodicity without intervening offsets. HD\,28471\,b exhibits a more eccentric orbit than the other planets, which may be due to dynamical interaction, or a result of RV variation from an as-yet-undetected 4\ts{th} planet interior to this compact system. The detected planets lie close to a resonant configuration, indicating that the system may retain features of its natal configuration, with convergent migration potentially responsible for evolving the planets onto such short-period orbits. 
%228 words
\end{abstract}

% Select between one and six entries from the list of approved keywords.
% Don't make up new ones.
\begin{keywords}
techniques: radial velocities -- planets and satellites: detection -- planetary systems -- stars: individual: HD\,28471
\end{keywords}

%%%%%%%%%%%%%%%%%%%%%%%%%%%%%%%%%%%%%%%%%%%%%%%%%%

%%%%%%%%%%%%%%%%% BODY OF PAPER %%%%%%%%%%%%%%%%%%

\section{Introduction}

Super-Earths and sub-Neptunes are found on orbits shorter than 100\,d in up to 50\% of planetary systems \citep{Leleu2024,Izidoro2017,Bean2021,Mulders2018}. Results from the \textit{Kepler} mission revolutionised our understanding of planetary systems \citep{Kepler}, with this relatively unknown class of planet being commonly found on orbits shorter than that of Mercury. 
 
\begin{figure}
    \centering
    \includegraphics[width=0.95\columnwidth]{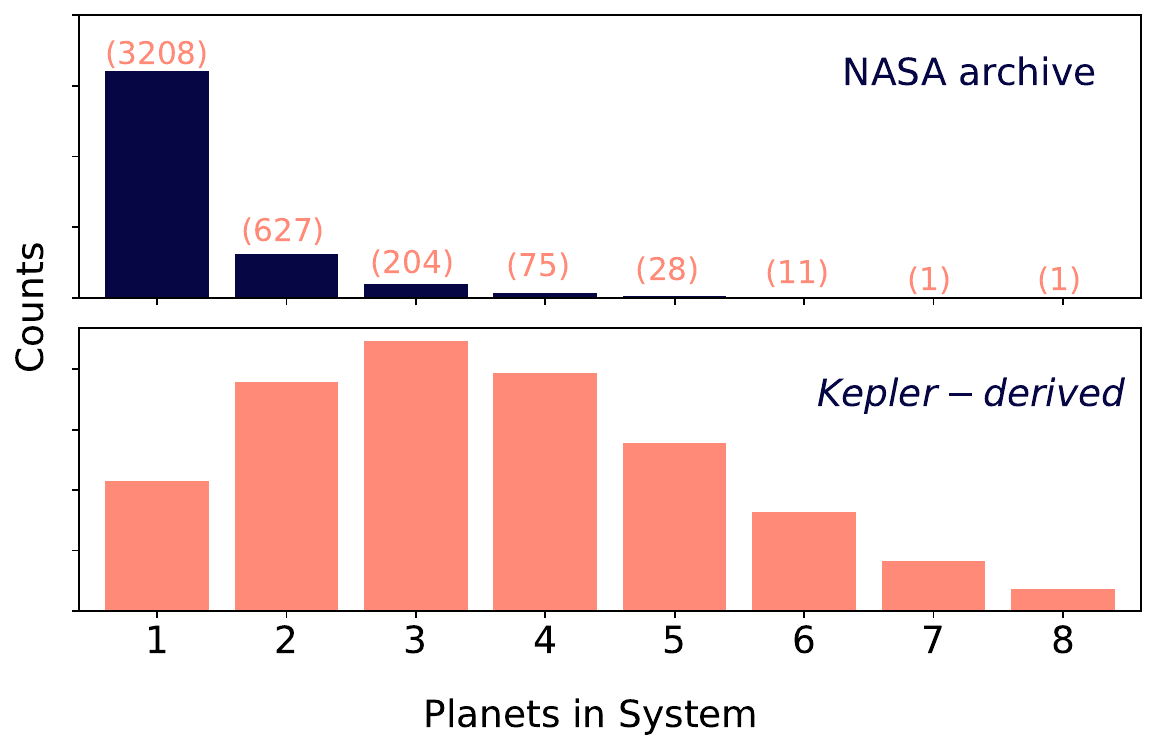}
    \caption{A comparison between the multiplicity of all currently known exoplanetary systems (NASA archive, 25 Mar 2024), and a prediction extrapolated from \textit{Kepler} demographics \citep{He2019}. Detections of multi-planet systems are far lower than predictions, primarily due to mutual inclinations between planets in transiting systems. In transit datasets, the observed excess of single-transit systems is known as the \textit{Kepler} dichotomy. Mutual inclinations are drawn from a broad range of Rayleigh distributions, with scale parameters optimised in the forward modelling process to make a simulated population reflect the observed population.}
    \label{fig:archive-vs-kepler}
\end{figure}

Forward models based on \textit{Kepler} statistics exploring period, mass, radii, and mutual inclination distributions predict far more multi-planet systems than 
there are present in our exoplanetary catalogues \citep*{He2019}, represented visually in Figure~\ref{fig:archive-vs-kepler}. Multi-planet systems should be common \citep{Batalha2013}, so it is important to identify the missing planets and fill in the demographics. Techniques like the radial velocity (RV) method, that are able to detect planets regardless of the mutual inclinations \citep{HaraFord2023}, 
can inform our picture of system architectures and our assumptions about mutual inclination distributions \citep{Mulders2018}. Accurate characterisation of multiplanet systems is needed to understand how they form, interact, evolve, and their eventual fate -- particularly as so many planets are on very close-in orbits.   

For multi-planet systems with measured masses and radii, \citet{peasinapod} found that the planet sizes were correlated between adjacent pairs. Where there are three or more planets, they also tended to have regular spacing between each pair. Smaller planets were seen to exist in increasingly compact systems, with the sizes and spacings a direct consequence of formation and subsequent dynamics \citep{peasinapod,Izidoro2017}. Period ratios between adjacent planets are more strongly correlated with planetary masses than radii \citep{Muresan2024}.
\citet{Adams2020} found that planets form with roughly equal-masses when the mass reservoir is not sufficient to form a massive giant planet. Probing system architectures therefore illuminates the evolutionary history, to compare with that of our own Solar System. 

\citet{Chance2024} identified that for planet pairs with one planet in the radius valley ($\sim1.8~\rm{R}_{\oplus}$), the radius ratio ($R_{\rm inner}/R_{\rm outer}$) peaks around 0.7-0.8. For a control sample of pairs with no valley planet, the radius ratio peaks at unity. Pairs with a valley planet therefore do not strictly follow the `peas-in-a-pod' pattern \citep{peasinapod}, as each planet is not as likely to be the same size as its neighbour.
Systems with unexpected characteristics may signal on-going evolution, as planets in the radius valley are expected to be losing mass via photo-evaporation \citep[e.g.][]{OwenWu2017,Fulton2017}. \citet*{Gilbert2025} performed an in-depth study of eccentricities for \textit{Kepler} systems, and found tantalising evidence that planets in the radius valley had elevated eccentricities in comparison to other super-Earth and sub-Neptunes. Eccentricity is very important in multi-planet systems from a formation viewpoint \citep[e.g.][]{Juric2008}, and particularly relevant for stability of closely-packed systems. Characterising sizes (either mass or radii), spacings between planets, and the eccentricities of the orbits, is therefore pivotal for understanding the nature of both individual systems and population demographics as a whole.

In this paper, we analyse radial velocity observations of HD\,28471, a low-activity G-type star harbouring a compact multi-planet system. The observations and star itself are described in Sections~\ref{sec:HD28471-obs} and~\ref{sec:HD28471-star}. A frequentist RV analysis is presented in Section~\ref{sec:recper}. Our Bayesian methodology is described in Section~\ref{sec:kima}, with results presented in Section~\ref{sec:HD28471-rvsol}. In light of these planetary detections, we search for transits in \textit{TESS} data for this target in Section~\ref{sec:HD28471-TESS}. Stellar activity is investigated in Section~\ref{sec:HD28471-activity}, and attempts to constrain the stellar rotation period are detailed in Section~\ref{sec:HD28471-Prot}. Finally, we discuss our results in Section~\ref{sec:discuss}, and conclude in Section~\ref{sec:conclusion}.

\section{Observations}\label{sec:HD28471-obs}

We use 122 spectra taken with the High Accuracy Radial velocity Planet Searcher (HARPS, \citealt{Harps}) at the ESO 3.6\,m telescope. Of these, 23 were taken as part of exoplanet searches from 2003 to 2013 (programs 072.C-0488(E) and 183.C-0972(A)). More intensive monitoring by the Dispersed Matter Planet Project (DMPP, \citealt{Haswell2020}) revealed periodic modulation. More details of the data are given in \citet[][PhD thesis]{StevensonThesis}. 

\subsection{HARPS regimes}\label{subsec:offsets}

Our measurements span two HARPS instrumental interventions: the 2015 fibre upgrade and the 2020 shutdown and warm-up. Both introduce RV offsets accounted for as a free parameter in our models\footnote{HARPS webpages recommend fitting for these offsets, i.e \url{https://www.eso.org/sci/facilities/lasilla/instruments/harps/news.html}}. We thus treat the data as if they were from three separate instruments. In Section~\ref{sec:HD28471-activity} and Appendix~\ref{app:bisector-offset}, we also assess if / how activity indicators are affected by these instrumental changes.

In 2015, the instrumental profile (IP) significantly changed during the fibre upgrade \citep[e.g.][]{Frensch2023}. The resulting RV offset is statistically significant, but has no constant value for all stars. There is evidence that the size of the RV offset is correlated with the width of stellar lines, where it could be expected that IP changes would interact with line width \citep{HarpsUpgrade}. From observing the response from RV stable stars, \citet{HarpsUpgrade} provided examples depicting RV offsets in the range of $~0-20$~m\,s$^{-1}$. \citet{Trifonov2020} found that G-type stars display a typical RV jump of $\sim11-14$~m\,s$^{-1}$, depending on reduction process used. These values provide an order of magnitude estimate for prior constraints to aid in RV modelling.

During the COVID-19 pandemic, HARPS warmed up. When cooled again, the instrument was re-calibrated, necessitating another RV offset in models \citep[as in][for example]{Stevenson2023-DMPP3}. We have studied the CCF FWHM from the HARPS \textsc{drs} and find a change in FWHM magnitude arises between data before and after the warm-up.

\subsection{RV reductions}
We re-reduced the RVs from the HARPS spectra. As detailed in Table~\ref{tab:rv_reductions}, we used three template matching alternatives to the CCF-based HARPS \textsc{drs} reductions (\citealt{HarpsTerra,SERVALpaper,S-BARTpaper}). Of the four reductions used here 
for HD\,28471, \textsc{s-bart} provides the best precision and the best consistency between the first two datasets and the third. Therefore we used the \textsc{s-bart} reductions in the RV analysis.

\begin{table}
    \centering
        \caption{Uncertainties on reduced RV values from different pipelines. The best precision is achieved using \textsc{s-bart}, so these were used for our analysis.}
    \begin{tabular}{lccc}
    \hline
         \multirow{2}{*}{Pipeline} & \multicolumn{3}{c}{RV uncertainty (m\,s$^{-1}$)} \\
         \cline{2-4}
          & Set~1  & Set~2 & Set~3  \\
         \hline
        $\textsc{drs}$ & $0.66$  & $0.51$ & $0.73$ \\
        $\textsc{serval}$ & $0.69$  & $0.79$ & $1.10$ \\
        $\textsc{harps-terra}$ & $0.67$ & $0.66$ & $0.88$ \\
        $\textsc{s-bart}$ & $0.54$ & $0.41$ & $0.58$\\
        \hline
    \end{tabular}
    \label{tab:rv_reductions}
\end{table}

\section{Stellar characterisation}\label{sec:HD28471-star}
We derived stellar parameters with the \textsc{PAWS} pipeline \citep{Freckelton2024} after co-adding all 122 spectra achieving a mean SNR per resolution element of $2084$.  \textsc{PAWS} uses the functionalities of \texttt{iSpec} \citep{blancocuaresma2014, blancocuaresma2019} to combine the Equivalent Widths (EW) and spectral synthesis methods, using \textsc{ATLAS9} model atmospheres \citep{kurucz2005}. 

The EW method was applied to the continuum-normalised spectrum using \textsc{WIDTH} \citep{width} to generate initial estimates of $T_{\rm eff}$, $\log g$, \text{[Fe/H]}, and $v_{\textrm{mic}}$. These estimates were used in the spectral synthesis method with the \textsc{spectrum} radiative transfer code and line list \citep{spectrum}. This yielded final values for $T_{\rm eff}$, $\log g$, \text{[Fe/H]}, $v_{\textrm{mic}}$, $v_{\textrm{mac}}$, and $v\sin i$. The reported uncertainties arise from the covariance matrix from least-squares fitting in the \texttt{iSpec} spectral synthesis. These reflect precision errors only, and do not account for model inaccuracies. To reflect the uncertainty in our models, we add an uncertainty of $100$\,K in quadrature to $T_{\rm eff}$.

The stellar mass, radius, and age come from the stellar atmospheric parameters derived above together with parallax and magnitudes with the \textsc{isochrones} package \citep{morton2015}, making use of the MESA Isochrones and Stellar Tracks (MIST, \citealt{dotter2016}). We followed \citet{mortier2020}: the spectroscopic $T_{\rm eff}$ and \text{[Fe/H]} are used with the Gaia DR3 parallax, 2MASS J, H, and K magnitudes \citep{skrutskie2006}, and AllWISE W1, W2, and W3 magnitudes \citep{cutri2013}. Our results are listed in Table~\ref{Tab:HD28471}.

The stellar rotation velocity is not precisely recovered through spectral analysis. The star is a slow rotator, with rotational broadening not adequately resolved with HARPS ($R\sim115000)$. Literature estimates suggest $v\sin{i} < 2$~km\,s$^{-1}$ (see Table~\ref{Tab:HD28471}).

Numerous studies have also reported stellar atmospheric parameters for HD\,28471. This bright star was included in HARPS GTO programmes, and as such, has thoroughly investigated properties. Our results in Table~\ref{Tab:HD28471} are consistent with those reported throughout the literature \citep[e.g.][]{2020AA..Hojjatpanah,DelgadoMena2021,Soubiran2022}

\begin{table}
	\centering
	\caption{HD\,28471 stellar parameters. Values are shown with corresponding $1\sigma$ uncertainties where appropriate. Parameters derived in this paper come from \textsc{PAWS} \citep{Freckelton2024}, \textsc{harps-terra} \citep{HarpsTerra}, and \textsc{actin2} \citep{ACTINsoftware}.
    }
	\label{Tab:HD28471}
	\begin{tabular}{lcc} % three columns, alignment for each
		\hline
		Parameter & Value & Reference\\
		\hline
        RA [hh mm ss] & 04 25 09.15  & \multirow{4}{*}{\citet{GaiaEDR3.2020}}  \\
        Dec [Deg mm ss] & -64 04 48.25  &   \\
		Parallax (mas) & $22.8701$ &   \\ 
		Distance (pc) & $43.7252$ &  \\ 
        Spectral type & G5V & \citet{1975mcts.book.....Houk} \\
		$V$ & $7.91$ & \multirow{2}{*}{\href{https://simbad.cds.unistra.fr/simbad/}{SIMBAD}} \\ %(apparent magnitude) 
		$B-V$ &  $0.65$ &  \\ %(apparent magnitude)
        \hline
        \multirow{2}{*}{$\log{R^{\prime}_{\rm HK}}$} & $-4.9948 \pm 0.0003$ & \textsc{harps-terra} \\
        & $-4.9819 \pm 0.0003$ & \textsc{actin2} \& \textsc{pyrhk} \\
        \hline
        %Rotation Velocities \\
		\multirow{3}{*}{$v\sin i$ (km s$^{-1}$)} & $1.6 \pm 1$ & \citet{2020AA..Hojjatpanah} \\
        & $1.846$ & \citet{2020AA...634A.136C} \\
        &  $1.737 \pm 0.208$ & \citet{2018..SPECIES.I} \\
  %       \hline
  %       $T_{\textrm{eff}}$ (K) & $5772 \pm 16 $ &  \multirow{8}{*}{\textsc{species}}\\
  %       \text{[Fe/H]} & $-0.04 \pm 0.015 $ &  \\
  %       $\log g$ (cm s$^{-2}$) & $4.36 \pm 0.02 $ & \\
  %       $v\sin i$ (km s$^{-1}$) & $3.78 \pm 0.08 $ &  \\
  %       $v_{\textrm{mac}}$ (km s$^{-1}$) & $3.18 \pm 0.08 $ &  \\
		% $R_{*} (\textrm{R}_{\odot})$ & $1.070 \pm 0.006 $ & \\
		% $M_{*} (\textrm{M}_{\odot})$ & $0.967 \pm 0.008 $ &  \\
		% Age (Gyr) & $7.34 \pm 0.50 $ & \\
        \hline
        $T_{\textrm{eff}}$ (K) & $5766 \pm 101$ &  \multirow{9}{*}{\textsc{PAWS}}\\
        \text{[Fe/H (dex)]} & $-0.04 \pm 0.03 $ &  \\
        $\log g$ (cm s$^{-2}$) & $4.38 \pm 0.06 $ & \\
        $v\sin i$ (km s$^{-1}$) & $<2.0$ &  \\
        $v_{\textrm{mic}}$ (km s$^{-1}$) & $1.07 \pm 0.11$ & \\
        $v_{\textrm{mac}}$ (km s$^{-1}$) & $4.21 \pm 0.16$ &  \\
		$R_{*} (\textrm{R}_{\odot})$ & $1.08 \pm 0.057$ & \\
		$M_{*} (\textrm{M}_{\odot})$ & $0.98 \pm  0.044$ &  \\
		Age (Gyr) & $6.56 \pm  1.94$ & \\
		\hline
        $P_{\textrm{rot}}$ (d) & $26.18 \pm 0.29$ & \multirow{2}{*}{\citetalias{MH2008}} \\ 
        Age (Gyr) & $4.56 \pm 0.19$ & \\
	\end{tabular}
\end{table}

HD\,28471 has consistently low activity. To confirm this we include two estimates of $\log R'_{\textrm{HK}}$ in Table~\ref{Tab:HD28471}. We use either the S-index from \textsc{harps-terra}, converted to the Mount Wilson system then to $R'_{\rm HK}$ \citep{1984..Noyes}; or the S-index output by \textsc{actin2}, converted with the \textsc{pyrhk} module \citep{ACTINsoftware,GomezdaSilva2021}. Additionally, \textsc{pyrhk} calculates a chromospheric rotation period based on the activity level and the $B-V$ colour, and a gyro-chronology age estimate. These are calculated using the empirically derived relations from \citet[][henceforth \citetalias{MH2008}]{MH2008}, and are also listed in Table~\ref{Tab:HD28471}. The activity-derived age estimate and that derived from isochrone fitting with \textsc{PAWS} are consistent to within $1\sigma$.

HD\,28471 exhibits sub-Solar chromospheric activity, as the Sun has $\log{R'_{\textrm{HK}}} = -4.91$ \citep{Burrows2024}. \citet{GomezdaSilva2021} found that the distribution of $\log{R'_{\textrm{HK}}}$ for G-type stars is bimodal, and that the low-activity mode peaks at $-4.95$~dex. With mean activity of $\log{R'_{\textrm{HK}}}\sim4.98$~dex, HD\,28471 is also less active than the majority of stars deemed ``inactive''. The epoch values of $\log{R'_{\textrm{HK}}}$ from either pipeline are fairly stable over the $\sim19$\,yr observational baseline ($-5.06<\log{R'_{\textrm{HK}}}<-4.94$), suggesting the star consistently remains in this low-activity regime.

However, the $\log{R'_{\textrm{HK}}}$ (and S-index) could reasonably be suppressed by circumstellar shrouds of material lost from hot, close-in planets. The DMPP hypothesis \citep{Haswell2020} suggests that when viewing a star through such a shroud, light will be noticeably absorbed in the resonance lines which can reduce the measured chromospheric emission to below the basal limit for main sequence stars. Despite not being below this basal limit, there is evidence for short period planets in HD\,28471, and it is not infeasible that mass-loss from any orbiting planets is reducing the $\log{R'_{\textrm{HK}}}$ \citep[e.g.][]{GAPS-51}. The true activity level may therefore be slightly higher than measured here.

\section{Recursive planet searches}\label{sec:recper}

As an initial search for planetary signals in the RV data, we employed the \textit{recursive periodogram} technique described by \citet{AT12}. The authors warned that parameter correlations and aliasing can hide signals in the residuals after subtracting a Keplerian model from RVs, and as such, developed a generalised approach to the classic least-squares periodogram that is optimised for multiplanet detection. The \textit{recursive periodogram} adjusts parameters of already-detected signals together with the signal under investigation. Each proposed periodicity is initially fitted with an assumed circular orbit, and several iterations at each test period are used to ensure that the fully--Keplerian solution converges. This approach has been used in other works such as \citet{Anglada2013} and \citet{Proxima-b}, where the recursive periodogram (and simultaneous adjustment of all other candidates) massively improves the significance of proposed signals, compared to a simple residual periodogram.

\subsection{Planetary candidates}
\begin{figure}
    \centering
    \includegraphics[width=1.0\columnwidth]{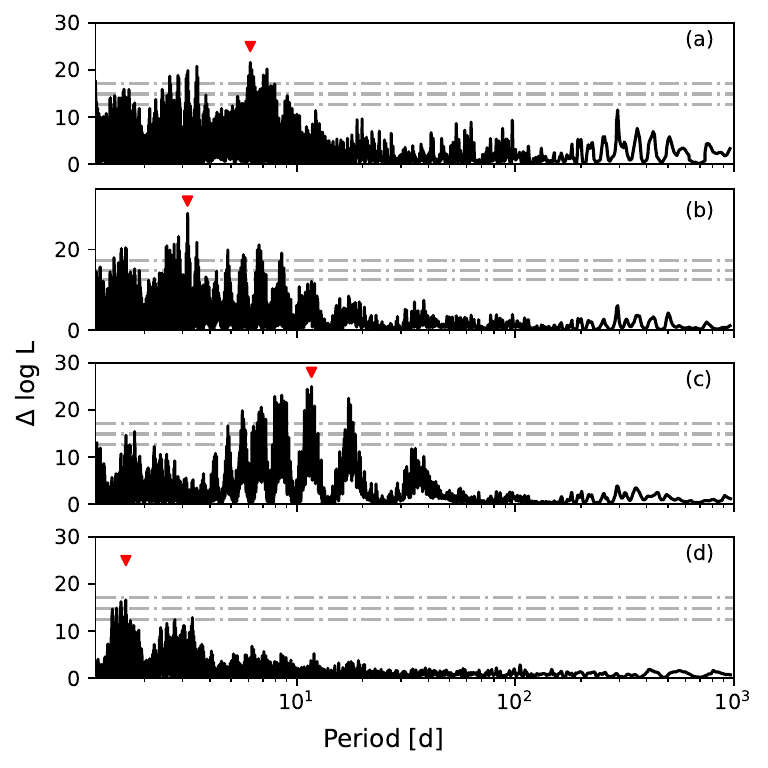}
    \caption{Log-likelihood periodograms for the HD\,28471 RV data. Periodograms are calculated recursively, after first fitting for the $\sim1500$\,d signal. Red markers indicated the periods corresponding to the maximum likelihood solutions. Dashed horizontal lines denote the 10, 1, and 0.1 per cent false alarm probability levels (in order of increasing power).}
    \label{fig:recper-periodograms}
\end{figure}

There are clear offsets between the three datasets and RV variability is observed within each dataset on different timescales. In addition to variability of a few m\,s$^{-1}$ in each observing run, there is a longer term trend of $\sim15$~m\,s$^{-1}$ in the post-upgrade dataset. The post-upgrade RVs (cf. Figure~\ref{fig:HD28471-Beta-rvsol}) correspond to the beginning of DMPP monitoring of this star \citep{Haswell2020}, where four observing runs were carried out in fairly close succession.

Periodograms identify a strong signal (false-alarm probability (FAP) equal to $7.48\times10^{-32}$ per cent) at $\sim1500$\,d as the first periodicity in the RV data. The recursive periodogram search continues from this point, identifying another 3 signals with analytic FAP substantially below the $0.1$~per~cent level commonly required for claiming planetary detection ($0.0049,0.000025,~\textrm{and}~0.0019$ per cent, respectively). The log-likelihood periodograms showing the detected periodicities are presented in Fig.~\ref{fig:recper-periodograms}. The strongest signals are detected at $6.12, 3.16, \&~11.68$\,d (shown in panels (a) to (c) of Fig.~\ref{fig:recper-periodograms}).

Searching for another periodicity, a signal is found at $\sim1.65$\,d with analytic FAP $\sim2$~per~cent (Fig.~\ref{fig:recper-periodograms}\,(d)). 
%The likelihood improves by $\ln{L}=16.65$, a Bayesian information criterion (BIC) improvement of $\Delta\textrm{BIC}=9.28$. The Bayes Factor (BF) in favour of including this planet in the model is BF~$=104$, 
This is below the threshold required to formally detect a planet, and ends the recursive periodogram search. Further observations may aid in increasing the detection significance of this periodicity in the residual RVs.

Though the recursive periodogram methodology has found several signals in our RV data, there can also be issues with this approach. The periodograms shown in Figure~\ref{fig:recper-periodograms} are heavily aliased and contain a number of fringes around the candidate planetary periods. The algorithm selects the most significant (lowest FAP, highest $\Delta\log{L}$) peak to fit a signal to, then adjusts this signal slightly when fitting subsequent Keplerians. For any incorrect selection, the code may fall into a solution set that represents a local minimum, not adequately exploring the full parameter space. Issues with this process were identified by works such as \citet{Barnes2023}, who found inconsistencies between frequentist and Bayesian approaches to the RV-fitting problem. Therefore, to confirm any planetary detections in HD\,28471 and fully explore the parameter space, we also analyse the RV data with a cutting-edge nested sampling approach.

\section{Bayesian RV analysis with \textsc{kima}}\label{sec:kima}

To search for Keplerian signals we used the \textsc{kima} package \citep{kima-joss,kima-ascl}. \textsc{kima} simultaneously fits Keplerian models to RV timeseries using Diffusive Nested Sampling (DNS) to explore the parameter space \citep{Brewer_DNS}. Such simultaneous modelling provides a far more robust estimation of the \emph{combination} of signals present within the data \citep{Barnes2023}, hence is ideally suited to compact multiplanet systems. The code samples from the posterior distributions for all parameters, and can calculate the marginalised likelihood (evidence) allowing for straightforward model comparison. The evidence ratio, or `Bayes factor' ($BF$) between models can be converted to probabilities or significances, in favour of one over the other.

\subsection{Trans-dimensional sampling}

\textsc{kima} not only models the parameters of Keplerian signals, but also infers \textit{how many} Keplerians are formally detected. The number of planets ($N_{p}$) is treated as a free parameter, and has an estimated posterior distribution just like any other parameter.

For our purposes, the model parameters in \textsc{kima} are 
\begin{equation}
    \theta = [N_{\textrm{p}}, \{P,k,e,\phi,\omega\}, s_{1}, s_{2}, s_{3},v_{\textrm{sys}}, \gamma_{12}, \gamma_{23}].
\end{equation} 
The Keplerian elements (period, semi-amplitude, eccentricity, mean anomaly, and argument of periastron) are denoted by $P,k,e,\phi,\textrm{and}~\omega$, respectively; these are repeated for each planet in the model. We have three white noise terms ($s_{1},s_{2},s_{3}$), for each of the three `instruments' (see Section~\ref{subsec:offsets}). The white noise is added in quadrature to the RV uncertainties
to account for stochastic astrophysical noise processes \citep{Aigrain2012,Luhn2020,HaraFord2023}. The systemic radial velocity of the reference data set is explored, where $v_{\rm sys}$, $\gamma_{12}$, and $\gamma_{23}$ are the systemic radial velocity and the two relative offsets between datasets, respectively. 

We used a Student's $t$-distribution \citep{student-t}, which robustly handles outliers in the data during model likelihood evaluations -- see \citet{Faria2022}, \citet{Standing2023}, and \citet[][PhD thesis]{StevensonThesis}.

\subsection{AMD-stability enforcement}\label{sec:kima-AMD}
Conditions for stability using the angular momentum deficit framework (AMD, \citealt{Laskar1997,Laskar2000,Laskar2017,PLB2017}) can be applied in \textsc{kima} to ensure each sample corresponds to a stable orbital configuration. This is discussed in more detail in \citet[][PhD thesis]{StevensonThesis}. This allows us to use less-restrictive eccentricity priors, reflecting the underlying statistics of currently known exoplanet orbits \citep[e.g.][]{Standing2022,Stevenson-ecc}.

\subsection{Bayes factors in \textsc{kima}}\label{sec:kima-BF}

The \textsc{kima} posteriors  include samples with a varying number of planets, as the model samples different $N_{\rm p}$ parameter combinations to effectively describe the RV variation. The ratios between number of samples for each planet, the posterior probability ratios, can be interpreted as the Bayes factor: the evidence ratio between competing models with different $N_{\rm p}$. For a derivation of this result, see \citet{lambert2018student} and \citet[][PhD thesis]{StevensonThesis}. 

Typically, $BF>150$ is required for `strong' evidence \citep{Trotta2008,Standing2022}. This can be assessed by comparing the number of samples, or posterior probability, for one value of $N_{\rm p}$ against another \citep{BrewerDonovan2015,Faria2016,Hara2022}. When there are no samples for a given $N_{\rm p}$, one instead quotes a lower limit on the Bayes Factor, $BF>N_{\rm p +1}/1$.   

Requiring that each successive planet that is ``confidently detected'' and added to the model satisfies this condition, we are considering false positive detections to be much worse than false negatives \citep{Faria2016}. This aligns with the Occam's razor principle, by preferring the least complex model. Often, the highest posterior probability lies at a larger $N_{\rm p}$ than we confidently detect.

\section{Radial Velocity Solution}\label{sec:HD28471-rvsol}

Our initial analysis performed in Section~\ref{sec:recper} suggests that the data contain both a long-period ($\sim1500$\,d) signal and a number of short-period ($P<15$\,d) Keplerians. Running \textsc{kima} with unrestricted period priors also identified two main groups of samples, at $P<20$\,d, and at $P\sim1500$\,d. This aligns with results from our recursive periodogram search, and informs the next stages of our analysis.

\subsection{Prior selection}

\subsubsection{Long-period signal}

In a blind $N_{\rm p}=1$ \textsc{kima} search for the aforementioned $\sim1500$\,d signal, a clear periodicity is detected in all samples, amounting to $BF\gg150$. We include this signal in all further analysis, adopting it as a ``known object'' (KO) with priors informed by the $N_{\rm p}=1$ runs (Table~\ref{tab:HD28471-KO-priors}). Including a KO effectively allows the use of non-continuous priors, to efficiently sample the long \textit{and} short-period Keplerian signals in this system \citep[e.g.][]{Standing2022,Baycroft2023a}. The KO RV signal will be referred to as the ``long-period signal'' or ``$1500$~d signal''; we will discuss whether it is planetary or a stellar activity artefact in Section~\ref{sec:HD28471-activity}.

\begin{table}
    \centering
    \setlength\tabcolsep{0.05\columnwidth}
    \begin{tabular}{lcc}
    \hline
    Parameter  & Prior  \\
    \hline
    Period, $P$ [d]    &  $\mathcal{U}\,[1300,\,1700]$   \\
    Semi-amplitude, $K$ [m\,s$^{-1}$]    &  $\mathcal{U}\,[5,\,20]$   \\
    Eccentricity, $e$     &  $\mathcal{K}\,(0.881,\,2.878)$   \\
    Longitude of periastron, $\omega$ [rad]    &    $\mathcal{U}\,[0,\,2\pi]$  \\
    Mean anomaly (at ref. time $t_{0}$), $M_{0}$ [rad]  &  $\mathcal{U}\,[0,\,2\pi]$   \\
    \hline
    Added white noise [m\,s$^{-1}$] & $\mathcal{MLU}\,(1; 16.0391\textsuperscript{\textdagger})$\\
    $v_{\rm sys}$ [m\,s$^{-1}$] & $\mathcal{U}$ (min RV, max RV) \\
    Offsets [m\,s$^{-1}$] & $\mathcal{U} (v_{\rm sys}\pm\textrm{RV range}$) \\
    \hline
    \end{tabular}
    \caption{`Known object' priors used to account for long-period variation in \textsc{kima} runs on HD\,28471 RVs, from Section~\ref{sec:HD28471-inner-planets} onwards. $\mathcal{MLU}$ is a modified log-uniform prior with a knee and upper limit. \textsuperscript{\textdagger}The upper limit is automatically determined based on the largest RV span of a single dataset. The priors result in well-sampled posteriors that are not restricted by the imposed limits.}
    \label{tab:HD28471-KO-priors}
\end{table}

\begin{figure*}
    \centering
    \includegraphics[width=0.90\linewidth]{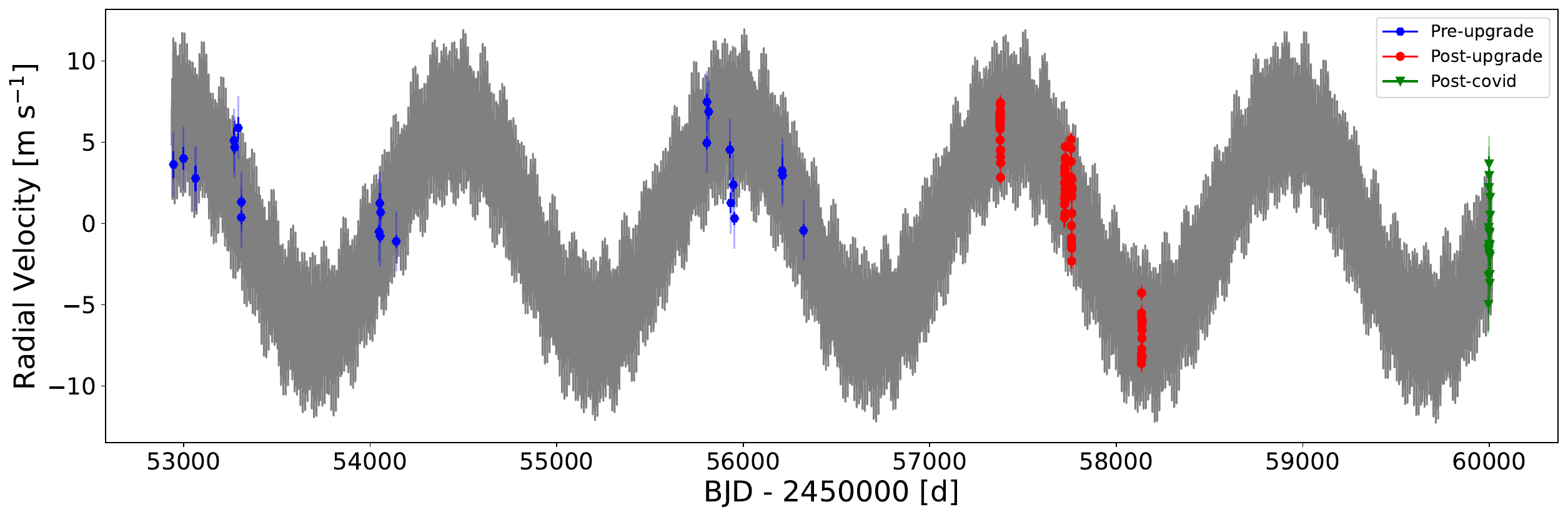}
    \caption{A diagram showing the HD\,28471 RVs, with the maximum likelihood $N_{\rm p}=3$ (+ KO) solution over-plotted (grey). The offsets have been accounted for between datasets, and these sets are denoted by different colours and symbols (as described in the figure legend).}
    \label{fig:HD28471-Beta-rvsol}
\end{figure*}

\subsubsection{Inner system}\label{sec:HD28471-inner-planets}

Analysis in Section~\ref{sec:recper} showed evidence in favour of multiple short-period signals. With the long-period signal accounted for (see above), priors to recover the interior planets could be specified, and the maximum number of Keplerians increased in \textsc{kima}, to place a constraint on the number of planets in the system and characterise their orbital parameters. To achieve sensible outputs and well-sampled posteriors for this inner system, we investigated the optimal priors.

Recursive periodogram analysis indicates that the period prior must extend beyond $\sim12$\,d. We set the upper bound at 15\,d, after experimentation revealed that longer periods would begin to include occasional alias periodicities. These arise because the poor phase coverage of the KO signal provides flexibility to fit other signals alongside. The offsets between the three datasets gave the freedom to account for the poor fit by placing a spurious RV signal at the longest period allowed by the prior.

The prior on semi-amplitude was limited to between $0.1$~m\,s$^{-1}$ and $4$~m\,s$^{-1}$. The lower limit is set to some small value, not too close to zero, so the sampler explores the $K$ posterior sufficiently 
\citep{Standing2022}. The upper limit is set above the level of scatter on the data after accounting for the long-period signal. Additionally, all Keplerian signals identified in Section~\ref{sec:recper} have $K<\sim2$~m\,s$^{-1}$.

We initially used the Beta distribution recommended by \citet{kipping2013} as an eccentricity prior, implemented by using a Kumaraswamy ($\mathcal{K}$) distribution \citep{Faria2022} with $\alpha=0.881$ and $\beta=0.2.878$ \citep[see][]{Stevenson-ecc}. However, this resulted in parameter splitting and degeneracies, likely due to the overwhelming preference for circular orbits (with the AMD stability criterion also naturally selecting against higher eccentricity samples). Further details on this splitting are given in Appendix~\ref{sec:9or11}. 

Recently, \citet{Stevenson-ecc} investigated the eccentricity distribution of the currently known RV-detected exoplanets. They found that the population is no longer best-described by the Beta distribution, and suggested a mixture model of Rayleigh and exponential ($\mathcal{RE}$) distributions as a possible alternative. The main difference to the Beta prior is that their $\mathcal{RE}$ probability density function peaks at a non-zero eccentricity, though decays to 0 beyond $e\sim0.1$. As it provides slightly less weight to a circular orbit (reflecting the exoplanet population), we employed this prior distribution in our analysis. The degeneracies have been all-but-entirely removed, vastly cleaning up the period posterior distributions.

In the subsequent \textsc{kima} simulations, we find that these priors (listed in Table~\ref{tab:HD28471-planet-priors}) are adequate, as the posterior samples do not pile-up at the imposed boundaries. Further discussion of the exploration of prior choices outlined in this subsection is given in \citet[][PhD thesis]{StevensonThesis}.

\begin{table}
%\begin{adjustbox}{width=0.95\textwidth}
    \centering
    \setlength\tabcolsep{0.02\columnwidth}
    \begin{tabular}{lcc}
    \hline
    Parameter  & Prior & Note (see text) \\
    \hline
    $N_{\rm p}$ & $\mathcal{U}\,[0,\,5]$ & uniform in \textsc{kima} \\
   $P$ [d]    & $\mathcal{U}\,[1,\,15]$ &  Signals beyond $\sim15$~d artefacts    \\
    $K$ [m\,s$^{-1}$]    &  $\mathcal{U}\,[0.1,\,4]$   & Scatter $<8$~m\,s$^{-1}$ \\
    $e$  &  $\mathcal{RE}\,(0.68,3.32,0.11)$  & see \citet{Stevenson-ecc}  \\
         %&  $\mathcal{RE}\,(0.50,4.57,0.09)$  & \texttt{EccentriciPy} prior generated for $<100$~d (`short period') \\
         %&  $\mathcal{U}\,(0,\,0.4)$  &  To see impact. Samples with Beta all $e<0.4$ \\
    $\omega$ [rad]    &    $\mathcal{U}\,[0,\,2\pi]$ & Free angular parameter \\
    $M_{0}$ [rad]  &  $\mathcal{U}\,[0,\,2\pi]$  & Free angular parameter  \\
    \hline
    \end{tabular}
    \caption{Planetary priors used in \textsc{kima} runs for HD\,28471 throughout this work. Offset and additive white noise priors are shared with those for the KO, described in Table~\ref{tab:HD28471-KO-priors}.}
    \label{tab:HD28471-planet-priors}
   % \end{adjustbox}
\end{table}

\subsection{Planet candidates}\label{sec:HD28471-inner-results}
 
\textsc{kima} allows us to constrain the number of planets in the system. We confidently detect three planets, with \textit{BF}\,$>\sim48000$. In our discussion, $N_{\rm p}$ refers to only the short-period planets. The KO is an additional signal.

The number of samples for each $N_{\rm p}$ is shown in Table~\ref{tab:HD28471-BFs}.
The Bayes factor (evidence ratio) is calculated for each solution by dividing the number of samples for each $N_{\rm p}$ by the number of samples for $N_{\rm p}$-1 (cf. Section~\ref{sec:kima-BF}). Working upwards, requiring $BF>150$ to accept a successive planet in the model, 3 planets are accepted with $BF>48827$ (there are no samples for $N_{\rm p}=2$). The 4-planet model provides a slightly disfavoured fit over the 3-planet model ($BF=0.87$). As such, the 3-planet solution is accepted here. 

Accordingly, we have derived results here for these candidate planet signals using samples where $N_{\rm p} = 3$ \emph{only}. For samples with $N_{\rm p}>3$, the same periodicities appear as in the $N_{\rm p} = 3$ samples. However, the $N_{\rm p} = 3$ Keplerian parameters will be modified by the addition of further signals, hence we use only samples with $N_{\rm p} = 3$.

\begin{table}
%\begin{adjustbox}{width=0.95\textwidth}
    \centering
    \setlength\tabcolsep{0.02\columnwidth}
    \begin{tabular}{ccc}
    \hline
    $N_{\rm p}$ (inner planets)  & Number of posterior samples & $BF$ \\
    \hline
    0 & 0 & / \\
    1 & 0 & / \\
    2 & 0 & / \\
    3 & 48827 & >48827 \\
    4 & 42262 & 0.87 \\
    5 & 5020 & 0.12 \\
    \hline
    \end{tabular}
    \caption{Bayes factors for the \textsc{kima} analysis reported in Section~\ref{sec:HD28471-inner-results}. The three-inner-planet solution is accepted with $BF_{32}>150$. The number of effective posterior samples is 96119, from 500,000 raw samples.}
    \label{tab:HD28471-BFs}
   % \end{adjustbox}
\end{table}

\begin{figure}
    \centering
    \includegraphics[width=0.99\linewidth]{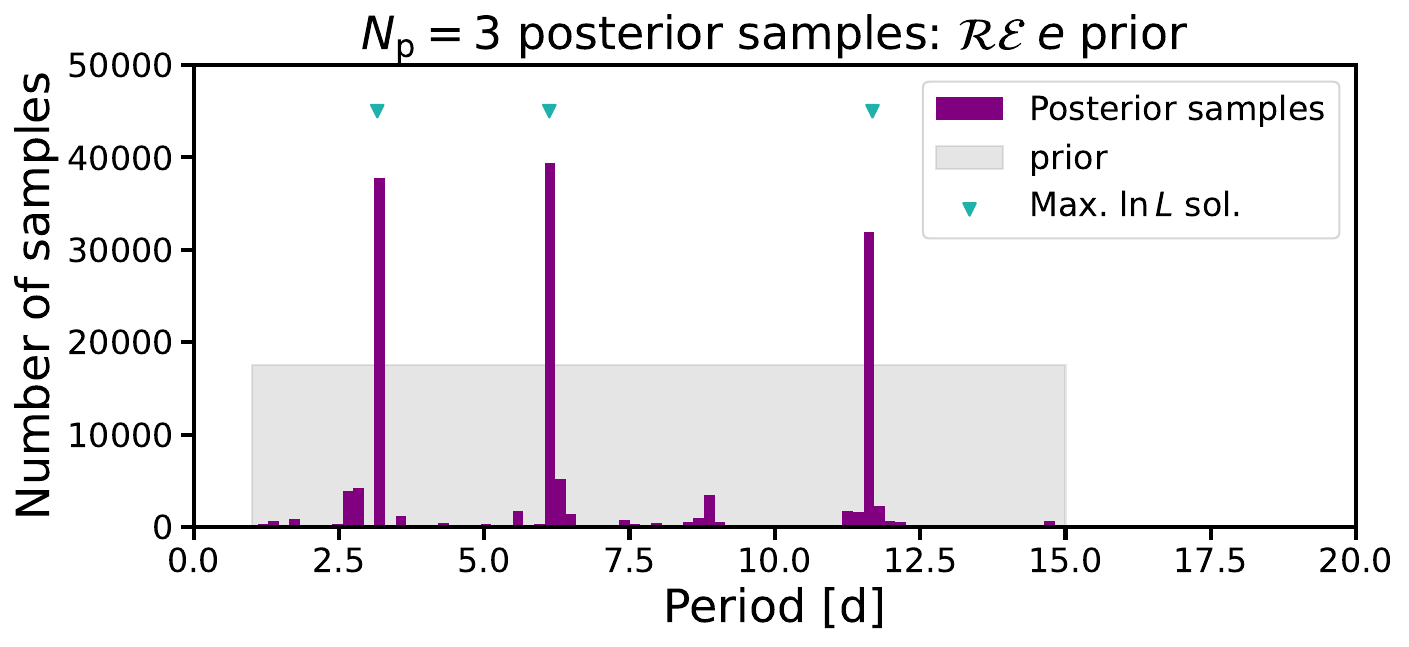}
    \caption{Period histogram for the $N_{\rm p}=3$ samples in our \textsc{kima} run. The three most prominent periods can easily be identified (also corresponding to the maximum likelihood solution), and match the periods detected with \textit{recursive periodogram} searches.}
    \label{fig:HD28471-RE-phist}
\end{figure}

The period posterior distribution (Fig.~\ref{fig:HD28471-RE-phist}) indicates the most probable periods found during the \textsc{kima} run. The three dominant periodicities are immediately apparent, at $P=3.16, 6.12,\&~11.68$. This mirrors the confidently detected inner planets from Section~\ref{sec:recper}, achieved with recursive periodogram analysis. For this particular RV dataset, the maximum likelihood and maximum \textit{a posteriori} solution match, providing consistent results between the frequentist and Bayesian approaches. The parameters and $1\sigma$ credible intervals for this 3-planet solution (plus the KO) are shown in Table~\ref{tab:HD28471-planet-params}.

\begin{figure}
    \centering
    \includegraphics[width=0.95\linewidth]{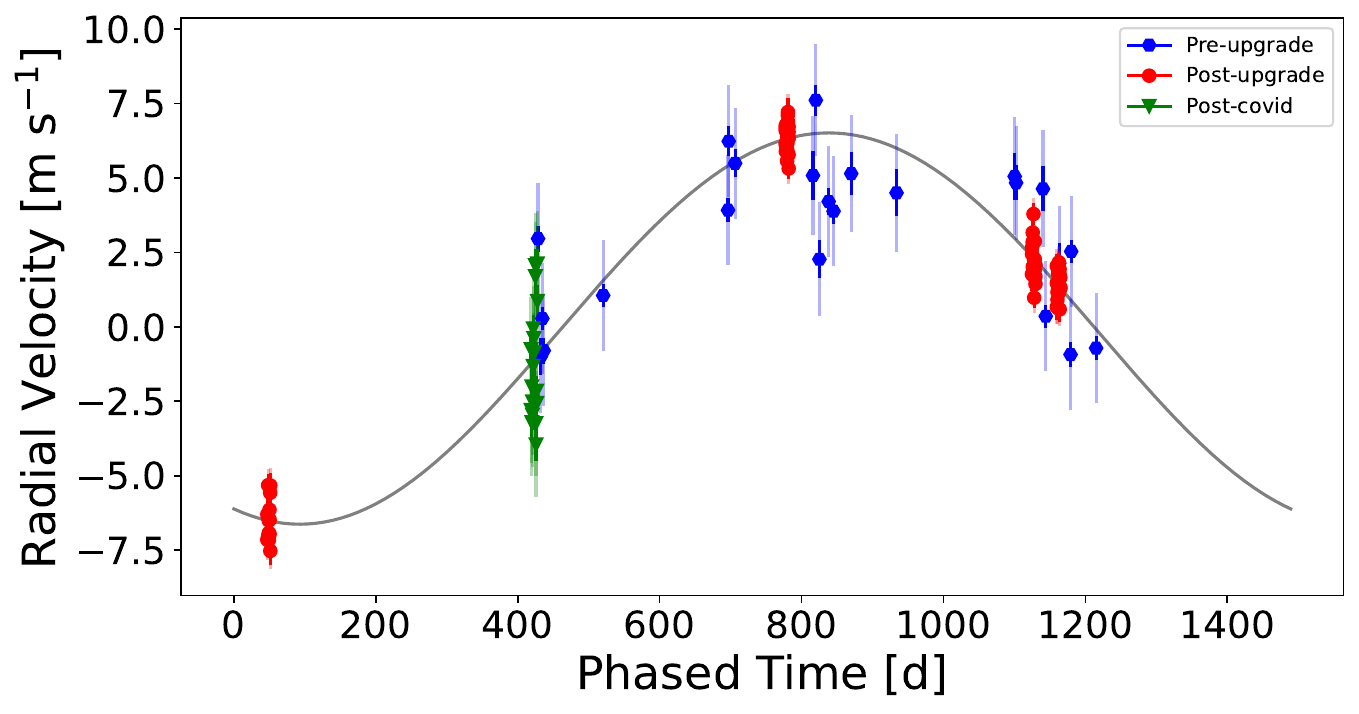}
    \includegraphics[width=0.95\linewidth]{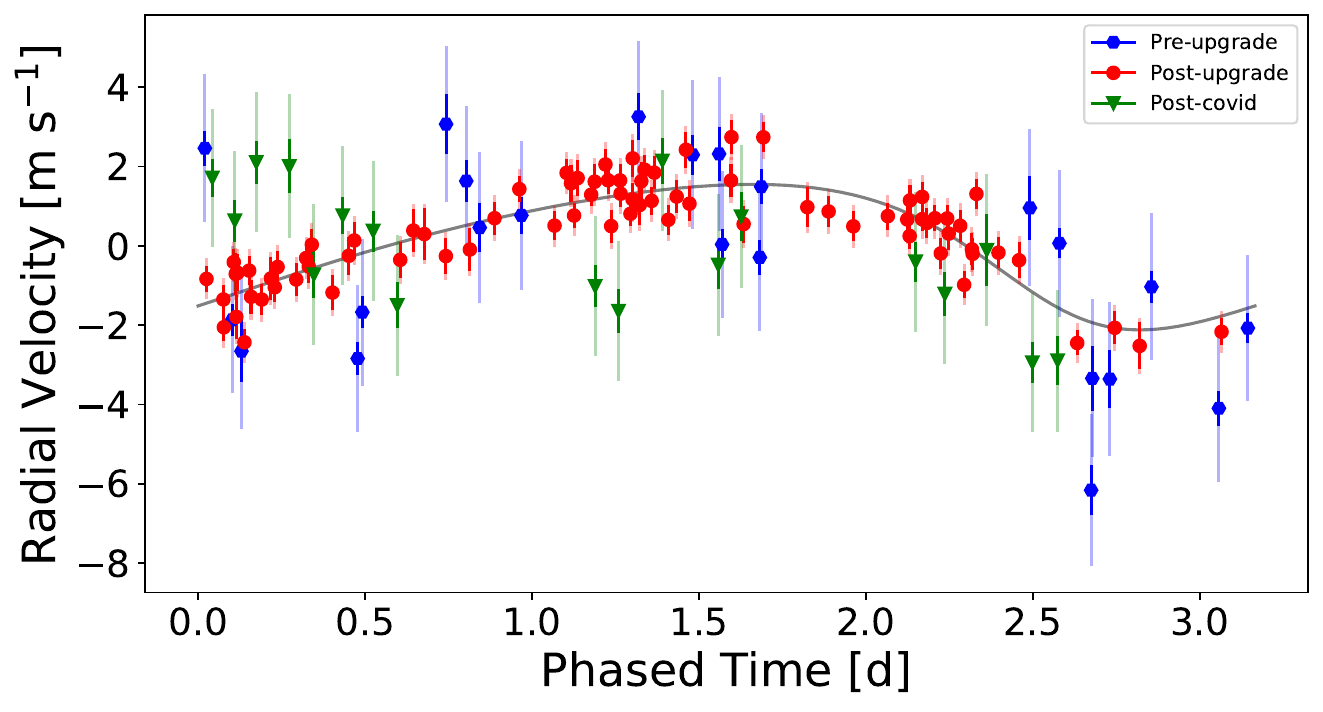}
    \includegraphics[width=0.95\linewidth]{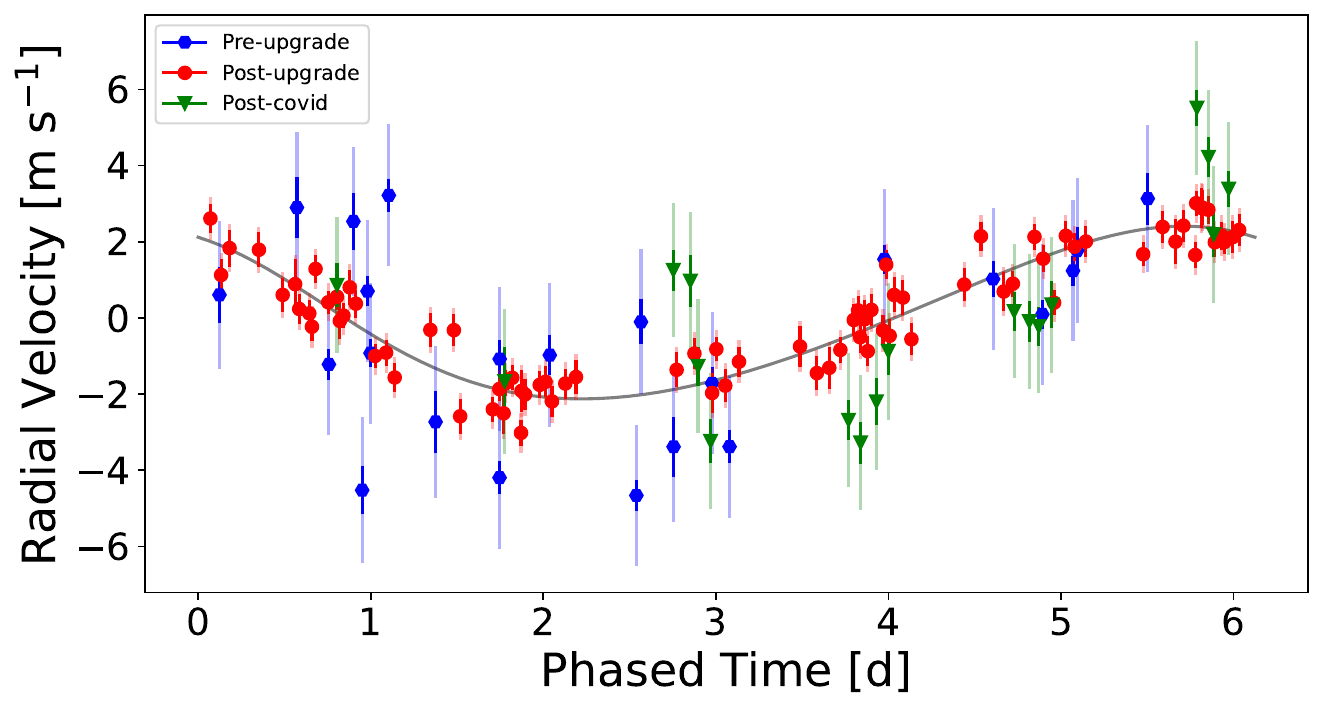}    
    \includegraphics[width=0.95\linewidth]{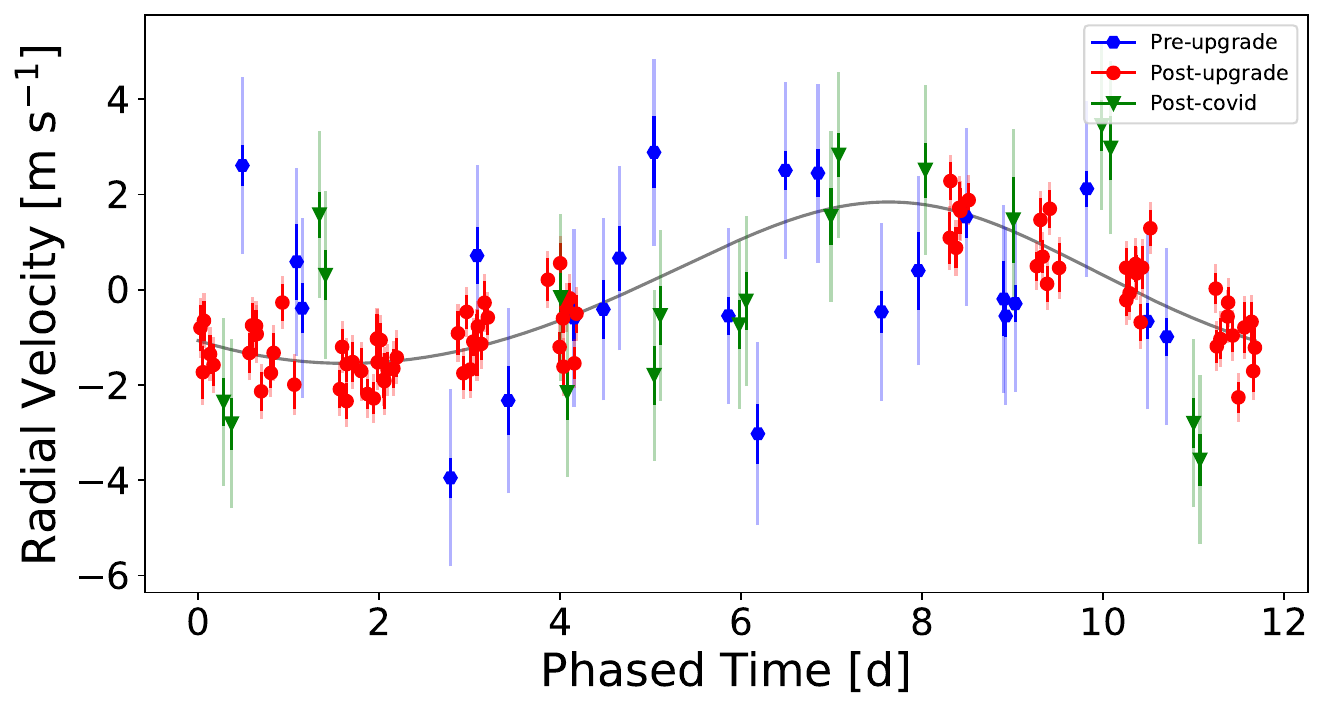}  
    \caption{Phase-folded RV plots for the favoured 3-planet (+ KO) solution in HD\,28471. \textbf{Top to bottom:} the long-period signal, then planets b, c, d, respectively. The parameters correspond to the maximum likelihood $N_{\rm p}=3$ solution in Table~\ref{tab:HD28471-planet-params}. Datasets are plotted in different colours and symbols, with additional white noise values added in quadrature to each set. The inflated error bar is plotted in a lighter shade of the same colour.}
    \label{fig:HD28471-Beta-folds}
\end{figure}

\begin{figure*}
    \centering
    \includegraphics[width=0.99\linewidth]{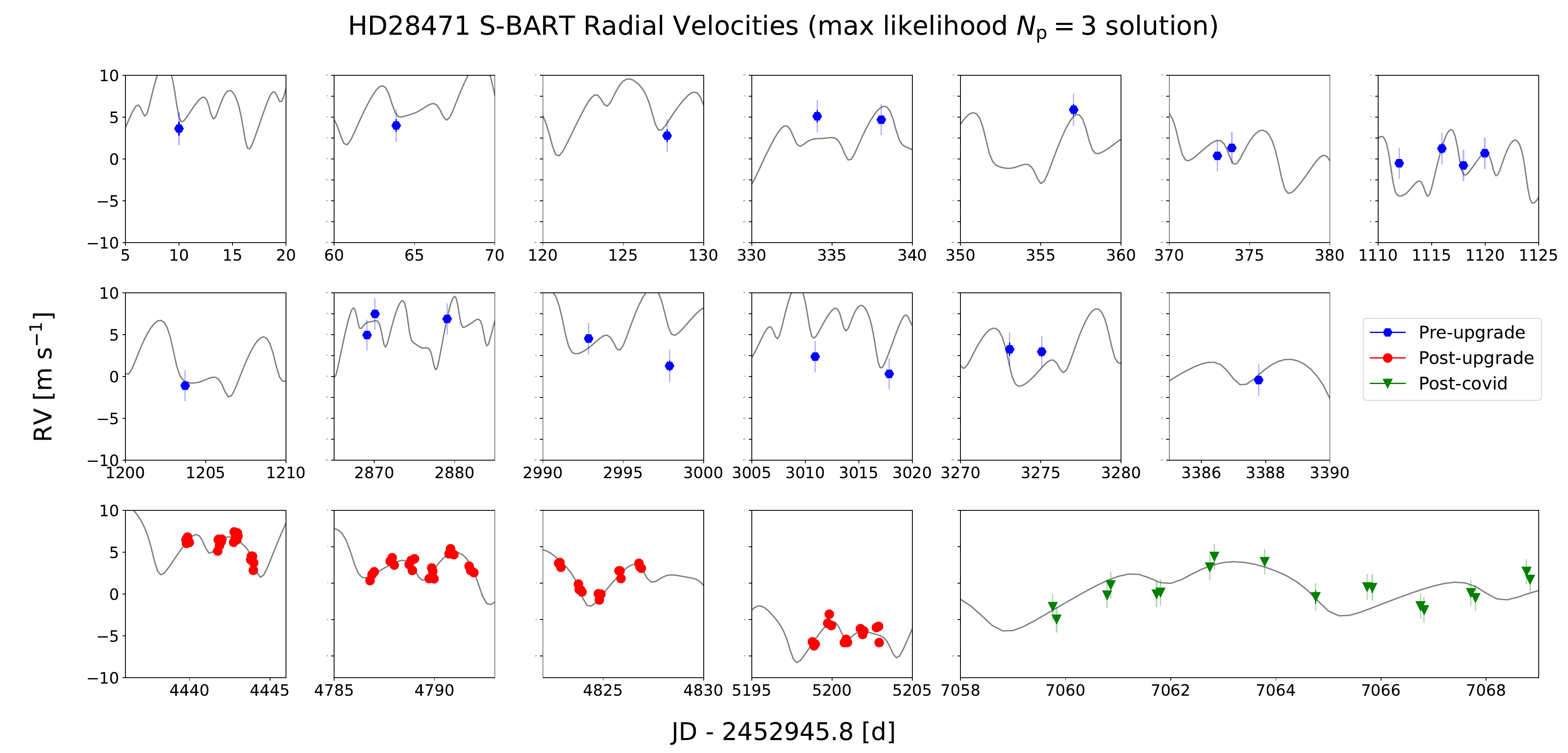}
    \caption{The HD\,28471 RVs split into epochs of closely-grouped observations, to assess the fit of the maximum likelihood $N_{\rm p}=3$ (+ KO) solution model. Error bars are extended in a lighter tone by adding the fitted white noise terms in quadrature. Colours and symbols correspond to different instrumental eras, described in the legend. Splitting panels in this manner gives a better view of the variation seen during densely sampled HARPS visitor mode observing runs (red and green).
    }
    \label{fig:HD28471-Beta-subplots}
\end{figure*}

To make phase-folded plots demonstrating the suitability of this solution, one needs to use a single sample. We selected the maximum-likelihood sample \citep{kima-joss,kima-ascl} for $N_{\rm p}=3$. This sample is included in square brackets in Table~\ref{tab:HD28471-planet-params}, and shown for the entire RV data-train in Figure~\ref{fig:HD28471-Beta-rvsol}. The phase-folded RV curves are shown in Figure~\ref{fig:HD28471-Beta-folds}, and each temporal cluster of observations is plotted in the panels of Figure~\ref{fig:HD28471-Beta-subplots}.

\renewcommand{\arraystretch}{1.8}
\begin{table*}
%\begin{adjustbox}{width=0.95\textwidth}
    \centering
    \setlength\tabcolsep{0.02\columnwidth}
    \begin{tabular}{lcccc}
    \hline
       Parameter & `Long period signal' & HD\,28471\,b  & HD\,28471\,c & HD\,28471\,d \\
    \hline
        $P$ (d) & $1495.57^{+12.26}_{-12.49}$ [1489.41] & $3.1649^{+0.0002}_{-0.0003}$ [3.1651] &  $6.1245^{+0.0856}_{-0.0009}$ [6.1236] & $11.6810^{+0.0042}_{-0.0055}$ [11.6783] \\
        $K$ (m s$^{-1}$)& $6.70^{+0.25}_{-0.17}$ [6.573]  & $1.679^{+0.180}_{-0.197}$ [1.836] & $2.045^{+0.190}_{-0.257}$ [2.267] &  $1.411^{+0.233}_{-0.218}$ [1.696] \\
        $e$ & $0.005^{+0.005}_{-0.004}$ [0.009] & $0.195^{+0.061}_{-0.073}$ [0.260] & $0.088^{+0.058}_{-0.054}$ [0.127] & $0.093^{+0.064}_{-0.058}$ [0.090] \\
        $\omega_{\textrm{0}}$ (rad) & $3.26^{+1.89}_{-1.85}$ [3.060] & $2.66^{+0.40}_{-0.45}$ [2.207] & $1.32^{+3.88}_{-0.78}$ [1.054] & $1.67^{+3.56}_{-1.08}$ [0.222] \\
        $M_{\textrm{0}}$ (rad) & $3.56^{+1.95}_{-2.56}$ [5.454] & $0.70^{+3.90}_{-0.43}$ [0.872] & $3.58^{+1.11}_{-1.11}$ [2.267] & $1.79^{+3.42}_{-1.23}$  [1.372] \\
        $M_\textrm{p}\sin{i}$ & $0.372^{+0.019}_{-0.016}~\mathrm{M}_{\textrm{Jup}}$\tnote{*}* & $3.72^{+0.40}_{-0.43}~\mathrm{M}_{\oplus}$ & $5.72^{+0.57}_{-0.72}~\mathrm{M}_{\oplus}$ & $4.91^{+0.82}_{-0.77}~\mathrm{M}_{\oplus}$\\
        $a_\textrm{p}$ (au) & $2.54^{+0.04}_{-0.04}$ & $0.042^{+0.001}_{-0.001}$ & $0.065^{+0.001}_{-0.001}$ & $0.100^{+0.002}_{-0.002}$\\
        $\gamma_{1}$ (m s$^{-1}$)  &  \multicolumn{4}{c} {$54126.93^{+0.53}_{-0.55}$ [54127.07]} \\
        $\gamma_{2}$ (m s$^{-1}$)  &  \multicolumn{4}{c} {$54143.05^{+0.22}_{-0.65}$ [54143.30]} \\
        $\gamma_{3}$ (m s$^{-1}$)  &  \multicolumn{4}{c} {$54148.04^{+1.32}_{-0.92}$ [54147.13]} \\
        $s_{1}$ (m s$^{-1}$)  &  \multicolumn{4}{c} {$2.25^{+0.65}_{-0.48}$ [1.88]} \\
        $s_{2}$ (m s$^{-1}$)  &  \multicolumn{4}{c} {$0.48^{+0.09}_{-0.09}$ [0.40]} \\
        $s_{3}$ (m s$^{-1}$)  &  \multicolumn{4}{c} {$1.37^{+0.54}_{-0.53}$ [1.68]} \\
        $N_{\textrm{obs}}$  &  \multicolumn{4}{c} {$122$ ($23+81+18$)} \\
        Baseline (d/yr) & \multicolumn{4}{c} {7059/19.3} \\
        $t_{0}$ (BJD) &  \multicolumn{4}{c} {$2452945.806$} \\
        $\ln{\mathcal{L}}$ &  \multicolumn{4}{c} {$-152.99$} \\
        Evidence, $\ln{Z}$ &  \multicolumn{4}{c} {$-240.28$} \\
        Information, $H$ (nats) &  \multicolumn{4}{c} {$76.47$} \\
         \hline
    \end{tabular}
    \begin{tablenotes}
  \item[*] *Mass of an orbiting body \emph{if} this signal is dynamical in nature.
  \end{tablenotes}
    \caption{
    Planetary parameters for the HD\,28471 system. Tabulated values are extracted from posterior samples of the preferred 3-planet (+ KO) solution, taking the median and uncertainties from \textsc{corner} quantiles. The maximum likelihood $N_{\rm p}=3$ solution is listed in square brackets, used to create the RV plots (Figs.~\ref{fig:HD28471-Beta-rvsol}, \ref{fig:HD28471-Beta-folds}, and~\ref{fig:HD28471-Beta-subplots}). The global parameters shared between the planets (e.g. absolute RV-offsets, $\gamma$, and white noise terms, $s$) and are presented in a single column in the latter half of the table.
    }
    \label{tab:HD28471-planet-params}
    %\end{adjustbox}
\end{table*}

\subsection{Another inner candidate}\label{sec:HD28471-4inner}

A solution with $N_{\rm p}=4$ has slightly lower but statistically indistinguishable evidence to the $N_{\rm p}=3$ solution. While clearly not preferred, there are $42262$ posterior samples with $N_{\rm p}=4$; they reveal possible parameters of additional planets that are not formally detected.
 
The $N_{\rm p}=4$ period posterior includes an additional strong peak near $\sim 1.65$\,d, present in almost every sample. The maximum log-likelihood for this solution is $\ln{\mathcal{L}}=-143.79$, an improvement of $\Delta\ln{\mathcal{L}}=9.2$ over the best-fitting $N_{\rm p}~\mathrm{(inner)}=3$ solution. 
 
Signs of this fourth inner planet were also seen in the frequentist approach, though had false-alarm probability below that required to accept a planetary detection. The periods are fully consistent, 
providing very tentative evidence of another possible signal. Further high-cadence RV observations may reveal this planet. Its estimated semi-amplitude is $\sim0.5$\,m\,s$^{-1}$, requiring high precision instruments such as ESPRESSO \citep{2013..ESPRESSO}.

\section{\textit{TESS} lightcurves}\label{sec:HD28471-TESS}

HD\,28471's proximity to the \textit{TESS} continuous viewing zone means it has been observed in 33 sectors.

\begin{figure}
    \centering
    \includegraphics[width=0.95\linewidth]{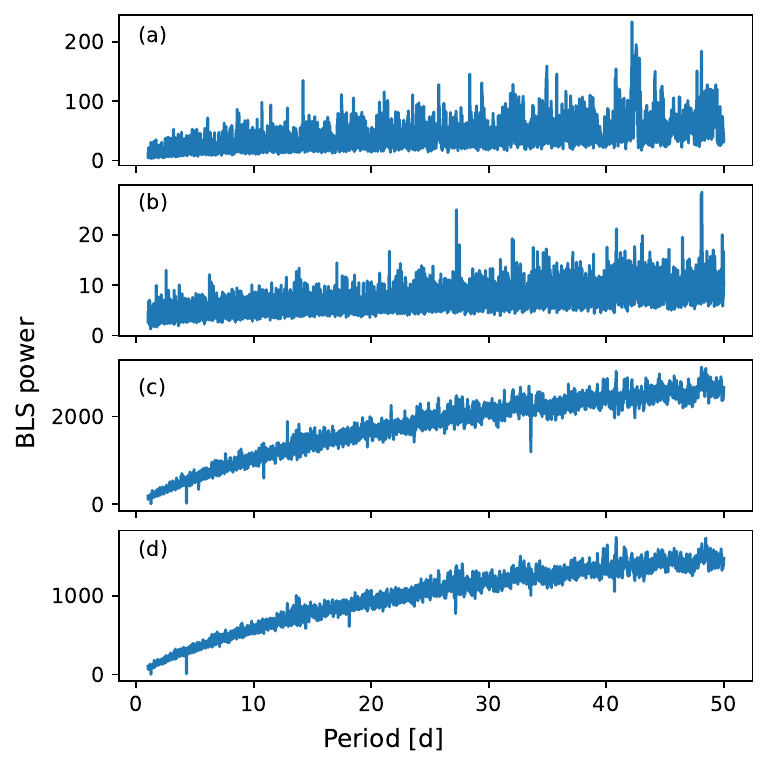}
    \caption{The BLS periodograms for HD\,28471 \textit{TESS} sectors. Lightcurves from each sector are stitched together (a,b,c,d), and flattened (b,d) and/or binned (c,d). The only prominent peaks above the noise level seem to be related to the \textit{TESS} sector baseline of $\sim27$\,d (see panel (b)) and aliases of this timespan.}
    \label{fig:TESS-BLS}
\end{figure}

We have searched these light-curves for any signs of transiting planets. We used BLS periodogram searches \citep{BLS} in the \textsc{lightkurve} package \citep{lightkurve}. There are no immediate signs of transiting planets in HD\,28471. BLS plots of the entire dataset, stitched together with the appropriate \textsc{lightkurve} function (Fig.~\ref{fig:TESS-BLS}), show no prominent peaks above the noise level, other than at periods close to $1\times$ and $2\times$ the \textit{TESS} sector length. We also tried binning the dataset (into 0.05\,d bins), and flattening the photometry time-series. The \texttt{.flatten()} function removes low frequency/long period variation by applying a Savitzky–Golay filter, effectively smoothing-out the data. This can help reduce stellar variability when searching for short-period transiting planets. 

We also performed BLS searches on the individual sectors. The highest-power period in each sector varies, and the periodograms are typically noisy and uninformative. Most of the BLS power is removed when flattening the light-curves, as instrumental discontinuities may have been contributing to what were identified as potential dimmings (hence periods are detected near $\sim27$\,d). The light-curves are visibly quiet in general, and consequently, bespoke techniques may be required to identify transit signals with depths close to the data noise levels \citep[as in, e.g.,][]{Jones2020}. The \textit{TESS} light-curves will be revisited to investigate the rotation period of HD\,28471 in Section~\ref{sec:HD28471-Prot}.

\section{Stellar activity}\label{sec:HD28471-activity}

We analyse the activity indicators to see if any of our RV signals may be due to activity. We computed periodograms for commonly used activity indicators, and assessed the correlations between them and the RVs. Offsets between datasets have been removed via median subtraction in cases where it was obviously needed (NaD emission), or is advised by \citet{HarpsUpgrade} and \citet{Trifonov2020}: BIS, CCF FWHM (and dLW by extension), CCF Contrast, and CCF Area. Correlations are assessed for each dataset independently, so that normalisation or RV offset determination does not impact any possible correlation between RV and indicator values.

\subsection{CCF indicators}

The CCF full-width at half-maximum (FWHM), contrast (depth) and area \citep{2019.Collier.Cameron} can be used to track changes in the mean spectral line profile. Before one can use the FWHM and contrast, instrumental drifts must be corrected \citep{2021MNRAS...Costes} and any offsets present in the time-series removed \citep[see Appendix~B in][]{Stevenson2023-DMPP3}.

\begin{figure}
    \centering
    \includegraphics[width=0.99\linewidth]{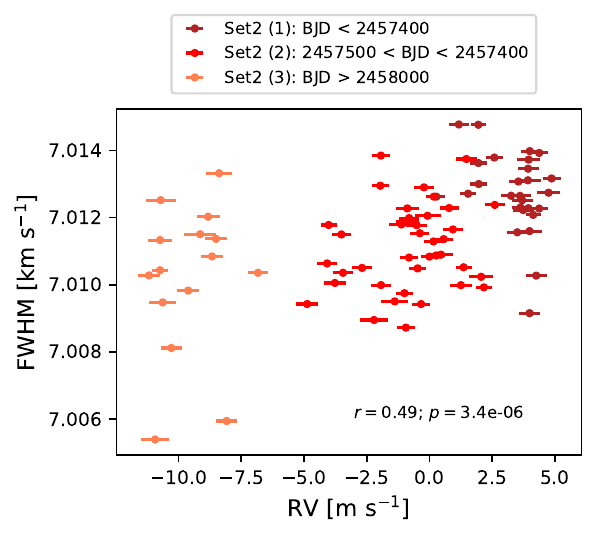}
    \caption{FWHM plotted against RV for set 2 only, splitting temporal clusters by colour shade. The `significant' correlation is detected due to FWHM changes between three observing runs, indicating that long-period activity processes may be the cause of the $\sim1500$\,d RV variation.}
    \label{fig:HD28471-FWHMset2only}
\end{figure}

Period analysis is performed with generalised Lomb-Scargle (GLS) periodograms, shown in Fig.~\ref{fig:HD28471-DRSindicator-GLS}. There are no significant periodicities detected in the CCF contrast and area periodograms, though there is a small peak at a period close to that of the long-period (KO) signal ($\sim1400-1500$\,d) in the FWHM periodogram. Inspecting the correlations between FWHM and RV (looking at each dataset individually) reveals that there are some `significant' ($p<0.05$) correlations (Fig.~\ref{fig:HD28471-DRSindicator-correlations}). Visually, this is not overly convincing. When comparing the red `post-upgrade' RVs shown in Figure~\ref{fig:HD28471-Beta-rvsol} to the FWHM data, the FWHM measurements are seen to follow a similar downwards trend (albeit with lower gradient), causing the detected correlation. This is also plotted in Figure~\ref{fig:HD28471-FWHMset2only}, this time splitting Set 2 into the three observing runs. The correlation arises from changes in the FWHM between these three groupings, indicating that the long-period RV signal may be of non-dynamical origin.

There may be some long-term activity modulation present on HD\,28471, with the change in FWHM tracking the global activity level -- in turn, affecting the RVs. It could also be an unrelated fortuitous correlation, with a separate cause. The post-upgrade data points are the sole driver behind the long-period signal, so it is worth being cautious over the interpretation of the signal. 

To investigate the activity interpretation of the long-period signal, we also assess indicators produced by the \textsc{serval} code. The chromatic index (CRX) and the differential line width (dLW) complement the CCF-based indicators, informing us of the chromaticity of the RVs (RV gradient as a function of wavelength) as well as providing a FWHM-analogue for measuring the line width when using a template matching reduction method \citep{SERVALpaper,SERVALascl}. 

The GLS periodograms for these indicators are shown in Figure~\ref{fig:HD28471-servalindicator-GLS}. Again, a forest of fringed peaks is observed, reminiscent of the window function between $\sim20$ and $\sim200$ days. The dLW periodogram has a significant peak near our KO signal at $1500$\,d, where FWHM did not (vertical dashed line). This measure of the width of the spectral lines is similar to FWHM, but results in a more significantly-detected periodicity in the 1000--2000\,d region. It is no surprise that some long-period variation is seen in dLW when it is seen in FWHM, as these indicators are, by design, usually strongly correlated \citep{Jeffers2022}.

Figure~\ref{fig:HD28471-servalindicator-correlations} shows the dLW is correlated with RV in a similar fashion to the FWHM-RV correlation (Fig.~\ref{fig:HD28471-DRSindicator-correlations}), as to be expected. The CRX data looks to be correlated with RV for Sets 1 and 3 (based on $p$-value $<0.05$), though is not correlated in the post-upgrade (red, Set 2) dataset. Set 2 contains the majority of the observations, so this indicates that some of the observed correlation may have been created by chance. There are no significant periodicities in the CRX periodogram, other than those at $\sim100$\,d, which are probably caused by sampling and the offsets. The CRX measurement can also be subject to telluric contamination, likely introducing spurious periodicities to the data.

A long period signal attributable to activity was also found in DMPP-3 \citep{Barnes2020,Stevenson2023-DMPP3}, where the $\sim 800\,$d signal is dynamically impossible due to the orbital configuration of the binary star. In DMPP-3, as in HD\,28471, the periodicity/correlation was only detected in the FWHM, with other indicators showing little evidence of activity. It is possible other long-period signals currently interpreted as planets may be of similar origin. Unlike the DMPP-3 FWHM signal, the FWHM periodogram of HD\,28471 does not have a significant (FAP $<10$~per~cent) peak at the questioned period, and there is no dynamical restriction to the existence of a planet on the candidate orbit. However, the dLW for HD\,28471 does exhibit a stronger signal, with false alarm level below $0.1$~per~cent, providing evidence that the signal may be caused by activity. Consequently, the long period signal has uncertain origin: more data is required (without additional offsets hampering the ability to detect long period variation) to truly confirm if the long period signal is caused by an orbiting body rather than stellar activity.

\subsubsection{Bisectors}\label{sub:lineprofiles}

\begin{figure}
    \centering
    \includegraphics[width=0.9\columnwidth]{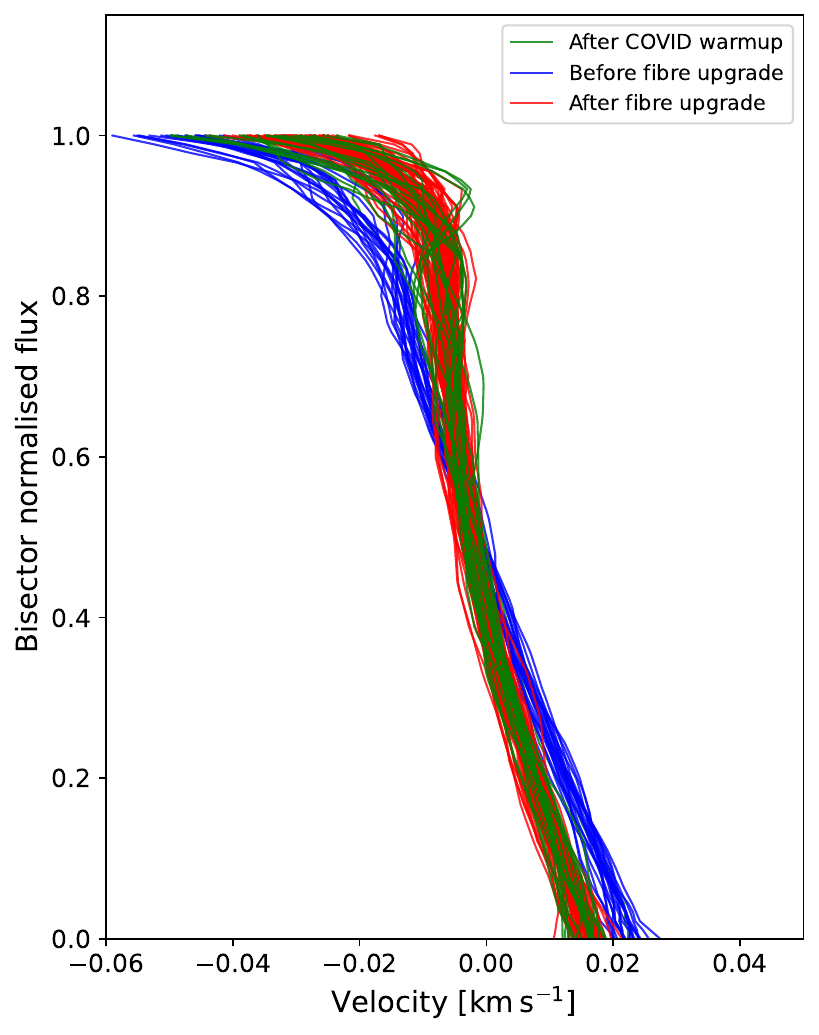}
    \caption{
    Line bisectors for HD\,28471. Data-sets are colour-coded, and all are horizontally shifted by the \textsc{DRS} RV value for each observation, to assess any changes in the shape. These curves are well-behaved within each observation set, with the majority of observations displaying a very similar shape.
    }
    \label{fig:HD28471-bisectors}
\end{figure}

Measuring changes in the line shape may not be very sensitive for HD\,28471 because the $v\sin{i}$ of the star is very low, making it difficult to resolve any activity-related velocity effects on the stellar surface. To assess any line shape variations, we plot the line bisectors in Fig.~\ref{fig:HD28471-bisectors}. The corrected CCF RVs for each observation have been subtracted off to bring the lines together. The bisectors appear to be very well behaved, although different sets of observations have subtly different slopes and overall shapes. Those recorded \textit{before} fibre upgrade can easily be visually separated from those taken \textit{after} the fibre upgrade. The change in instrumental profile \citep{HarpsUpgrade} therefore does seem to affect the bisectors too -- and this is discussed further in Appendix~\ref{app:bisector-offset}.

\subsection{Spectral lines}
As discussed in Section~\ref{sec:HD28471-star}, the S-index appears stable for this low-activity star. Nevertheless, we computed periodograms and assessed correlations with RV for this and other line emission strengths commonly used as activity proxies (Sodium doublet and Hydrogen $\alpha$). The GLS periodograms for all three are plotted in Figure~\ref{fig:HD28471-ACTINindicator-GLS}, where the indicators have been derived from the HARPS spectra with the \textsc{actin2} software. S-index has been corrected to the Mount Wilson (MW) system with the \texttt{calc\_smw} funciton of the \textsc{pyrhk} module \citep{ACTINsoftware,GomezdaSilva2021}

The periodograms of these indicators (Figure~\ref{fig:HD28471-ACTINindicator-GLS}) display a number of significant peaks, though the periodograms are extremely noisy. The forest of peaks observed in the $10$--$100$\,d range are similar to that seen in the window function (included in Fig.~\ref{fig:HD28471-DRSindicator-GLS}). The Na D periodogram shows peaks above the 0.1\% FAP level beyond $1000$\,d, close to the period of the long-period ($\sim1500$~d) signal (vertical dashed line). 

Assessing the correlations for these elemental line indicators, Figure~\ref{fig:HD28471-ACTINindicator-correlations} shows only a single panel with any significantly detected correlation: the post-upgrade Na\,D emission (Pearson's $r=0.65,~p=3.9\times10^{-11}$). It appears that during this era (red points), as in FWHM, there is a slight downward trend in the Na\,D indicator. The detected period is not exact as there is only one dataset constraining the variation, and it resembles a trend rather than sinusoidal variation. The cause for enhanced sodium emission is unclear, as other indicators (such as S-index and, by extension, $\log{R'_{\mathrm{HK}}}$) show no variation on similar timescales. The fact that S-index is far more commonly used as a magnetic activity proxy, but does not support the activity interpretation here, causes suspicion in the observed Na\,D changes. 

The Na\,D lines typically show smaller chromospheric effects than the S-index. They probe a region slightly higher in the chromosphere than the calcium lines do, so the differing levels of observed activity could be related to differing spatial origins \citep{GomezdaSilva2011}. The Na\,D wavelength region is commonly affected by telluric contamination. The \textsc{actin2} calculation of these indices may not be effectively dealing with telluric lines.

However, possibly we should also see varying levels of calcium emission, but the S-index is suppressed by attenuation in circumstellar material \citep{Staab2017,Haswell2020,Staab2020,GAPS-51}. It is important to explore numerous activity indicators.
 
Figures~\ref{fig:HD28471-ACTINindicator-GLS} and~\ref{fig:HD28471-ACTINindicator-correlations}, give no cause to attribute any short period variation in the RVs to stellar activity. The inner-system planets remain consistent with a dynamical interpretation.

\section{Rotation Period Searches}\label{sec:HD28471-Prot}

\subsection{Spectroscopic}

We attempted to determine the rotation period of HD\,28471, by inspecting periodograms of activity indicators presented in the previous section. When considering Figures~\ref{fig:HD28471-DRSindicator-GLS},~\ref{fig:HD28471-servalindicator-GLS}, and~\ref{fig:HD28471-ACTINindicator-GLS}, the most prominent peaks are observed in H$\alpha$, Na\,D, S-index, CRX, and dLW. Amongst significant alias fringing, these periodograms share features in the range of 15-40~days, with varying levels of relative power. The periodograms are very noisy, so perhaps rotational variation is not detected at all -- the window function included in these figures shows substantial power over the same interval. 

One can also calculate the rotation period from the $v\sin{i}$ of HD\,28471. All estimates (Table~\ref{Tab:HD28471}) concur that the projected rotation velocity of this star is below the resolution limit of HARPS. Taking $v\sin{i}\sim2$~km\,s$^{-1}$ as an upper limit, and using $P_{\textrm{rot}} = 2\pi R\,\sin{i}/~v\sin{i}$, we calculate a predicted rotation period of $P_{\textrm{rot}}/\sin{i} > 27.32$~d. This is consistent with the increased periodogram power observed in some of the activity indicators, and is very similar to the rotation period estimate based on the $\log{R'_{\rm HK}}$ activity level \citepalias[][see Table~\ref{Tab:HD28471}]{MH2008}. The estimated value confirms that HD\,28471 is a slow rotator, but does not accurately constrain any specific value. The estimate is too tenuous to inform hyperparameters used in GP regression, and is one of the reasons the technique has not been employed in this work (besides intrinsically very low activity, and poor temporal coverage outside 4 densely-sampled observing runs).

\citet{Yu2024} also investigated the rotation period of HD\,28471 as one of 268 targets in their sample, but their activity model did not adequately describe any indicator variation. This could partly be due to the chosen indicators: S-index and BIS. In our study, these indicators provide little information on periodic signals or activity correlation with the RVs.

The rotation period of HD\,28471 is intrinsically difficult to measure. When stars rotate slowly, the active region lifetime becomes comparable to the rotation period \citep{Yu2024}. Consequently, there is no periodic variation as the regions do not persist over multiple rotations, and therefore do not contribute to identifiable periodic fluctuations. 

\subsection{Photometric}

We also attempted to use photometry from \textit{TESS} to determine the rotation period of HD\,28471. Despite spectroscopic measurements being more sensitive to active region modulation than photometry for low-activity stars \citep{Yu2024}, there is a wealth of \textit{TESS} data available for this target. Running GLS period searches on the light-curves can be a useful way to identify any activity-related changes in brightness.

\begin{figure*}
    \centering
    \includegraphics[width=0.8\linewidth]{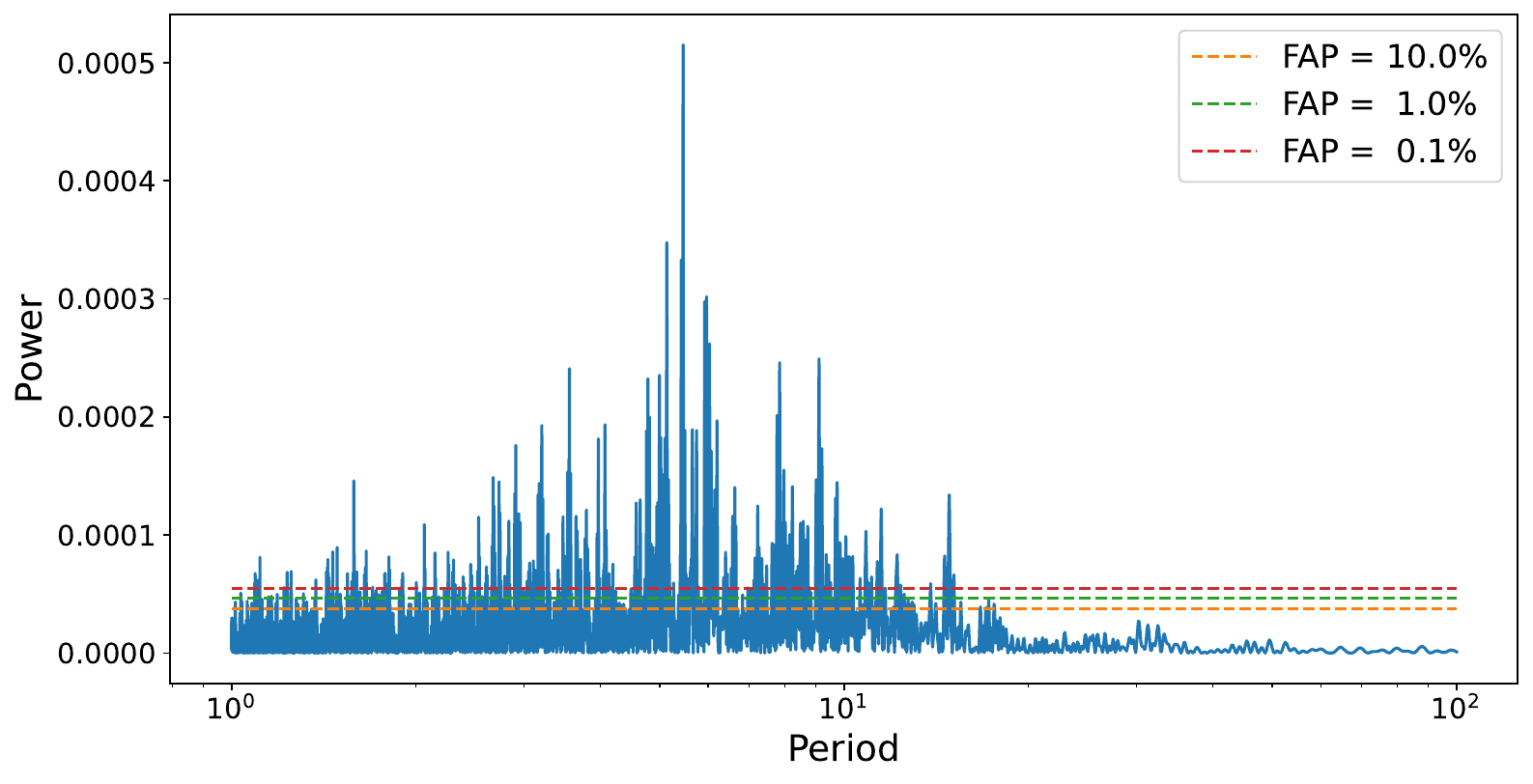}
    \caption{The GLS periodogram of 33 normalised \textit{TESS} sectors. The strongest peak is seen at $5.46$\,d, with a forest of `significant' peaks on either sides of this period. The 10, 1, and 0.1\% false-alarm levels are indicated by the dashed horizontal lines.}
    \label{fig:TESS-GLS}
\end{figure*}

The PDCSAP light-curves have been normalised from sector-to-sector to account for instrumental changes in the flux level, and converted into parts-per-million (ppm) flux units. The GLS periodogram (shown in Figure~\ref{fig:TESS-GLS}) identifies a strongest peak at $5.46$\,d. Many sub-peaks and fringes are found at significant FAP ($<0.1$\%) for periods below $\sim15$\,d. There is very little power at longer periods. Most of these peaks may be related to noise in the data, as minimal variation is observed in the light-curves. The highest-power period, at $5.46$\,d, has best-fit amplitude of $9.6 \pm 0.6$\,ppm -- far below the the weighted RMS level of the photometric dataset ($299$\,ppm). 

The best-fitting period identified varies in each individual sector, though upon inspection, most periodograms contain power in the $5$--$6$\,d region. This signal therefore does seem to persist over the baseline. It may be related to a spacecraft systematic, where a spurious periodicity is not fully corrected by pipeline de-trending.

In the preceding part of this section, we predicted that (based on the $v\sin{i}$) the stellar rotation period is longer than 27~d. A period detection around 5 days is far too short to be consistent with estimated rotation speeds, and would broaden the spectral lines enough to enable a consistently measured $v\sin{i}$ with spectral analysis techniques. A G-type star of this vintage ($\sim4-6$~Gyr; Table~\ref{Tab:HD28471}) is unlikely to spin this quickly \citep[e.g.][]{Skumanich1972,SBarnes2003,GalletBouvier2013}.

The strong $5.46$\,d detection may be a harmonic of the true rotation period, a common source of confusion when recovering rotation periods with periodograms \citep{Clarke2003,Vanderburg2016,Barnes-moments,Yu2024}. As each \textit{TESS} sector is of similar length to the lower limit on the rotation period, this may be confounding the period search.

\section{Discussion}\label{sec:discuss}

\subsection{Long period signal: activity}
The nature of the long-period signal ($P\sim1500$\,d) in HD\,28471 is one of the main points of contention arising from this work. There is a definite signal present in the RVs, evidenced by periodograms and the \textsc{kima} runs finding a strongly-detected long-period signal. This prompted us to account for the signal with a known object (KO) fit in subsequent runs. 

However, our investigations of stellar activity in Section~\ref{sec:HD28471-activity} indicate that the signal may be caused by variations in the star itself. Correlations between RV and FWHM, Na\,D, and dLW show that some activity indicators display a trend similar to the long-period signal. Although the origin could be unconnected, this has made the signal seem suspicious. 

This is very similar to a signal in the DMPP-3 system that was characterised as activity, which was the only remaining possibility after any dynamical origin was ruled out through N-body orbital simulations \citep{Stevenson2023-DMPP3}. As with DMPP-3, we have attempted to account for the variation with a simple fit, using the \textsc{kima} KO functionality here. Despite being uncertain about the signal nature, this remains the best way to remove the RV variation over such a long, poorly-sampled time-span. In the remainder of this section, we discuss any possible causes for such a signal, were it to be confirmed as activity. 

Although there is no clear activity modulation due to rotation in HD\,28471, this does not mean there is no activity -- active regions may decay\footnote{The lifetime of active regions is approximately $\sim50$\,d for lower-MS stars \citep{Donahue1997}. Small active regions on the Sun ($<20$ sq. deg.) have lifetime shorter than the solar rotation period \citep{Donahue1997}. For a slowly rotating star with low intrinsic activity (and relatively small active regions), the ARs may decay on a timescale much shorter than $P_{\rm rot}$.} before completing rotations \citep{Yu2024}, our observations may not be sensitive to the period, or the amount of the disc covered by active regions may inherently be very low ($\log{R'}_{\rm HK}$ quantifies this as a very inactive star). At each observation, the activity indicators will record a snapshot of the magnetic activity level. Over time, this may change depending on the global activity level of the star, modulated by the stellar cycle. 

Stellar cycles, like the Solar cycle, modulate activity on long timescales ($P\gg100$\,d). The number of active regions changes, which can affect activity indicators and RVs. Active regions can change the flux, and also inhibit the local convective motions. Where active regions suppress the convection, a red-shift is introduced when compared to the surrounding disc \citep{2018A&A...Bauer}. This changes both the RVs and activity indicators that relate to width and shape of the CCF: disc-integrated light will include regions with shifted lines, slightly broadening the average CCF and changing the shape \citep{2020A&A...Cretignier}. This is one of the main contributions to activity seen on low mass stars \citep{2021MNRAS...Costes,2020A&A...Cretignier,2021..Florian..CB} and is also a dominant part of the Sun's variability \citep{2010A&A...512A..39..Meunier}. The inhibition of convective blueshift is dependent on the global activity level \citep{2017A&A...597A..52Meunier}, tracing the stellar cycle. The amplitude of this signal ($14$~m~s$^{-1}$) is the correct order of magnitude, close to that seen for the Sun (11~m~s$^{-1}$). However, the $\sim4$~yr period is much shorter than that of the Sun's 11~yr cycle. This changes from star to star, though cycle lengths as short as $\sim1000$~d are still routinely discussed \citep[e.g.][]{2022MNRAS.514.2259Sairam,2016A&A...595A..12S-M}.

\subsection{Long period signal: planet}

Though it seems likely the long-period signal is due to stellar activity, if it is a planet it will affect the inner system. \citet{BryanLee2024} found that metal-rich ([Fe/H]>0) super-Earth hosts are more likely than field stars to also harbour a giant planet. This makes sense as protoplanetary disks with enough mass to provide dust to create a system of interior planets, would also have enough mass to nucleate a heavy core in the outer system \citep{ChachanLee2023}. Interpreted dynamically, the long period signal would be a $0.37\pm\sim0.02$~M$_{\rm Jup}$ planet, which could dynamically excite the inner system of super-Earths.

\subsection{Inner system}\label{sec:HD28471-discuss-innersys}

The analysis in this work has allowed identification of 3 confidently-detected ($BF>150$) inner planets in the HD\,28471 system, with signs of a potential 4\ts{th} planet candidate found in both frequentist and Bayesian approaches. Additional high-precision RV observations would help finalise the architecture and aid in resolving the true nature of the long-period signal. 

The $\sim3$, $\sim6$, and $\sim12$\,d, are close to a 1:2:4 resonant chain configuration. Analysing \textit{Kepler} detections, \citet{Fabrycky2014} found that there was an excess of planet pairs near first-order resonances, as the observed period distribution peaked within a few percent of these period ratios (e.g. $\sim1.505-1.520$ for the 3:2 resonance). 

The period ratio condition for pairs of (nearly) resonant planets can be expressed as
\begin{equation}
    \Delta_{\rm MMR} = \Bigg|\frac{P_{\rm outer}}{P_{\rm inner}} - \frac{k+q}{k}\Bigg| < 0.05,
\end{equation}
\noindent for $q=1,~k\in[1,2,3,4,5]$ or $q=2,~k\in[3,5]$ \citep{Leleu2024}. Non-resonant planets are defined such that $\Delta_{\rm MMR}>0.05$. The inner planet pairs in HD\,28471 exhibit $\Delta_{\rm MMR}\sim0.06-0.09$, and may either be close to resonant, or part of the excess of pairs just outside resonance noted by \citet{Fabrycky2014}.

The ``\textit{breaking the chains}'' model \citep{Izidoro2017,Bean2021} suggests that close-packed planetary systems naturally form in co-planar resonant chains, due to planetary migration within the disc and associated torques inflicted on the protoplanets. \citet{Leleu2024} found that surviving resonant systems are less-dense than their non-resonant counterparts, indicating that they retain their natal architectures as the planets have not lost their primordial atmospheres due to planet-planet collisions at some point in the system's history. Planets could however also enter MMR when migrating inwards, through convergent orbital migration \citep[e.g.][]{HaddenPayne2020}, without retaining the natal configuration

If the planets of HD\,28471 do exist in a close-to-resonant chain, it could be that it either retains the natal architecture, or has been captured into the current configuration through MMR migration \citep{Izidoro2017,Leleu2024,HaddenPayne2020,2022..Roisin}. It is therefore imperative to obtain further observations of HD\,28471 to precisely measure the period ratios between planet pairs in the system.

A near-resonant scenario for HD\,28471 may also help to identify any low-mass planets that are currently undetected in the RV data, but exist as part of a larger MMR chain. The detection of a resonant planetary system in HD\,110067 allowed additional planet periods to be probed with targeted observations and subsequently confirmed \citep{Luque2023}, and the same may be possible in this system. Consider that the planet on $\sim3.16$\,d orbit is not the start of the chain: one may predict that (using a range of periods for this planet) an interior planet would exist on a $\sim1.6$~d orbit. This is extremely similar to what is seen in Sections~\ref{sec:recper} and~\ref{sec:HD28471-4inner}, when including an additional (formally undetected) planet into our models. It may be that the HD\,28471 inner planets truly exist as a close-to-resonant quadruplet of super-Earths and sub-Neptunes. Measuring accurate and precise period ratios may help infer this planet's presence, even if it is too light to be readily detected in the RVs. 

On the other hand, there may not be an 4\ts{th} inner planet, and any potential RV variation comes from the fact that HD\,28471\,b appears to have a slightly eccentric orbit (see Figure~\ref{fig:HD28471-compecc} and increased eccentricity when compared with the prior). However, two planets in a 2:1 period configuration can appear in RV as an eccentric solution at the period of the outer planet \citep{Anglada2010,Wittenmyer2013,Boisvert2018,Wittenmyer2019}, and \textsc{kima} often finds eccentric signals when this is the case \citep[tested in][PhD thesis]{StandingThesis}

\subsection{Equilibrium Temperature}

To assess the impact the star may have on the planets, we have calculated the equilibrium temperatures for the inner HD\,28471 system (in a similar manner to \citealt{Barnes2023}) and tabulate the results for Bond Albedos corresponding to Mercury, Solar System gas giants and Venus in Table~\ref{tab:HD28471-Teq}. The planets are all very warm (as one would expect), but the innermost planet is especially hot. For non-Venusian albedo, it would have equilibrium temperature consistent with liquid magma on Earth, 1100--1500\,K \citep[][and references therein]{Staab2020,Barnes2023}. This planet may have been stripped of its atmosphere, and heated to such a level that the surface is melting and beginning to outgas volatiles from within the rock. Were a 4\ts{th} inner planet with $P\sim1.6$\,d to be detected, it would have an even higher estimated equilibrium temperature, in the range of 1200--1700\,K. Such an extreme environment could give rise to radiation--powered mass loss and facilitate the creation of a circumstellar shroud in the inner regions of this system.

\begin{table}
    \centering
    \begin{tabular}{l|c|c|c}
    \hline
    \multirow{2}{*}{\hfil Bond Albedo} & \multicolumn{3}{c}{$T_{\rm eq}$ [K]} \\
    \cline{2-4} 
     & HD\,28471\,b & HD\,28471\,c & HD\,28471\,d \\
    \hline
    0.088    & 1377 & 1066 & 945  \\
    0.36  & 1261 & 976 & 865  \\
    0.76   & 986  & 764 & 677 \\
    \hline
    \end{tabular}
    \caption{Equilibrium temperature estimates of the HD\,28471 inner planets using a variety of Bond albedos, informed by those observed in the Solar System \citep{Barnes2023}.}
    \label{tab:HD28471-Teq}
\end{table}

\section{Conclusions}\label{sec:conclusion}

\begin{enumerate}
    \item We have analysed a set of 122 HARPS RVs for the star HD\,28471, to discover any exoplanets it hosts. Using diffusive nested sampling and a model that includes the number of Keplerians as a free parameter, we find a favoured solution characterising the orbits of 3 close-in planets, with a long-period signal also present. This matches the solution obtained from a recursive periodogram search, adding further credibility to the planetary detections.
    \item Tentative signs of an additional interior planet are found, though do not pass the either the FAP or \textit{BF} thresholds required to confirm the detection. Due to the compact, low-amplitude nature of this system, and the 4\ts{th} inner planet having an estimated semi-amplitude below the precision of HARPS, ESPRESSO observations would be needed to reduce the ambiguity. Ultra-precise RVs will facilitate the search for an additional short-period planet, and refine the solutions to arrive at a finalised architecture of the system.
    \item BLS periodograms have revealed no indication of transits for HD\,28471, despite the wealth of available \textit{TESS} sectors (33). The detected planets may not lie in a transiting orientation, but could also be too small to be recovered from the noise without bespoke analyses. 
    \item Inspecting activity indicator periodograms and RV correlations has led us to suspect that the long-period signal in HD\,28471 may be an artefact of stellar activity. No variation is present on short timescales, though the $\sim1500$\,d modulation may in reality be a stellar activity cycle. Correlation is seen for RVs vs FWHM, dLW, and Na\,D in Dataset~2, which contains the main body of the observations. Data spanning a longer timebase, with no intervening offsets, is needed to fully classify this signal as either stellar activity or a Saturn-mass giant planet.
    \item Our quest to find the rotation period of HD\,28471 has been unsuccessful, with activity indicators not providing useful constraints. We have identified a very low-amplitude sinusoidal signal in the \textit{TESS} photometry, at a period wholly inconsistent with the rotation period based on $v\sin{i}$ estimates. This is similar to the work of \citet{Yu2024}, who were unable to determine $P_{\rm rot}$ for HD\,28471 with Gaussian process regression on BIS and S-index activity indicators. 
\end{enumerate}

\section*{Acknowledgements}

These results were based on observations made with the European Southern Observatory (ESO) 3.6\,m telescope and HARPS, under ESO programme IDs: 072.C-0488(E); 096.C-0876(A); 098.C-0269(A); 098.C-0269(B); 0100.C-0836(A); 110.248C.001; and 183.C-0972(A). \\

Appendices~\ref{app:bisector-offset} and~\ref{app:dLW-trend} were based on archival observations made with the ESO 3.6\,m telescope and HARPS, under ESO programme IDs:
60.A-9036(A);
60.A-9109(A);
60.A-9709(G);
072.C-0488(E);
072.C-0513(D);
073.C-0784(B);
074.C-0012(B);
074.C-0364(A);
075.C-0332(A);
076.C-0878(A);
077.C-0530(A);
078.C-0833(A);
079.C-0046(A);
079.C-0681(A);
081.D-0065(D);
083.C-1001(A);
084.C-0229(A);
085.C-0019(A);
085.C-0318(A);
086.C-0230(A);
087.c-0831(A);
087.C-0990(A);
089.C-0050(A);
089.C-0415(B);
089.C-0497(A);
089.C-0732(A);
090.C-0849(A);
091.C-0034(A);
091.C-0936(A);
092.C-0579(A);
092.C-0721(A);
093.C-0062(A);
093.C-0409(A);
094.C-0797(A);
095.C-0040(A);
095.C-0551(A);
096.C-0053(A);
096.C-0460(A);
096.C-0499(A);
097.C-0021(A);
099.C-0458(A);
0100.C-0487(A);
0102.C-0525(A)
0102.C-0558(A);
0102.C-0584(A);
0103.C-0206(A);
0103.C-0432(A);
105.20AK.002;
105.20PH.001;
106.21SE.002;
0106.C-1067(A);
0108.C-0110(B);
0109.C-0134(A);
0109.C-0356(A);
109.239.001;
0110.C-4159(A);
183.C-0972(A);
192.C-0852(A);
192.C-0852(G);
196.C-1006(A);
and 198.C-0836(A).
\\

We thank the anonymous referee for their feedback and suggestions, significantly improving the quality and clarity of this manuscript. ATS and ZOBR were supported by Science and Technology Facilities Council (STFC) studentships. ATS gratefully acknowledges funding from the School of Physical Sciences at the Open University, to facilitate the completion of this project.
CAH and JRB are supported by grant ST/T000295/1, and ST/X001164/1 from STFC. JKB is supported by an STFC Ernest Rutherford Fellowship (grant ST/T004479/1). MRS acknowledges support from the European Space Agency as an ESA Research Fellow. AVF acknowledges the support of the Institute of Physics through the Bell Burnell Graduate Scholarship Fund.
\\

Stellar atmospheric parameters were determined using \textsc{PAWS}, accessible openly from \url{https://github.com/alixviolet/PAWS}. The authors would like to thank Annelies Mortier for her expertise and guidance on the use of the \textsc{isochrones} package. \\

Radial velocity data were analysed with the \textsc{kima} package, freely accessible online and described in \citet{kima-joss}. The authors wish to thank J. P. Faria for sharing his invaluable expertise and explanations of the \textsc{kima} software. \\

This research has made use of the SIMBAD data base, operated at CDS, Strasbourg, France. This paper includes data collected by the \textit{TESS} mission. Funding for the \textit{TESS} mission is provided by the NASA's Science Mission Directorate. This research has made use of the \href{https://exoplanetarchive.ipac.caltech.edu/}{NASA Exoplanet Archive}, which is operated by the California Institute of Technology, under contract with the National Aeronautics and Space Administration under the Exoplanet Exploration Program.\\

The following \textsc{python} modules have also been used in this work: \textsc{astropy}, \textsc{corner}, \textsc{numpy}, \textsc{scipy}, \textsc{lightkurve}, \textsc{matplotlib}, \textsc{pandas}, and \textsc{pyastronomy}.\\

Data collection involved in this publication would have not been possible without travel grants from the Royal Astronomical Society (RAS), allowing postgraduate students the opportunity to get in-person experience of observational astronomy that had previously been hindered by the COVID-19 pandemic.

%%%%%%%%%%%%%%%%%%%%%%%%%%%%%%%%%%%%%%%%%%%%%%%%%%
\section*{Data Availability}

%The inclusion of a Data Availability Statement is a requirement for articles published in MNRAS. Data Availability Statements provide a standardised format for readers to understand the availability of data underlying the research results described in the article. The statement may refer to original data generated in the course of the study or to third-party data analysed in the article. The statement should describe and provide means of access, where possible, by linking to the data or providing the required accession numbers for the relevant databases or DOIs.

RVs, ancillary data, and posterior sample files are included as journal supplementary information online.

%%%%%%%%%%%%%%%%%%%% REFERENCES %%%%%%%%%%%%%%%%%%

% The best way to enter references is to use BibTeX:

\bibliographystyle{mnras}
\bibliography{refs} % if your bibtex file is called example.bib

% Alternatively you could enter them by hand, like this:
% This method is tedious and prone to error if you have lots of references
%\begin{thebibliography}{99}
%\bibitem[\protect\citeauthoryear{Author}{2012}]{Author2012}
%Author A.~N., 2013, Journal of Improbable Astronomy, 1, 1
%\bibitem[\protect\citeauthoryear{Others}{2013}]{Others2013}
%Others S., 2012, Journal of Interesting Stuff, 17, 198
%\end{thebibliography}

%%%%%%%%%%%%%%%%%%%%%%%%%%%%%%%%%%%%%%%%%%%%%%%%%%

%%%%%%%%%%%%%%%%% APPENDICES %%%%%%%%%%%%%%%%%%%%%

\appendix

\section{Alternate eccentricity priors}\label{sec:9or11}
In Section~\ref{sec:HD28471-rvsol} we presented results using an eccentricity prior that was recently derived to match the known RV exoplanet population \citep{Stevenson-ecc}. This resulted in a cleaned-up period posterior, allowing accurate selection of the three detected inner planet signals. In this Appendix, we highlight some of the issues that appeared during our use the Beta distribution as an eccentricity prior \citep{kipping2013}, the standard for the field.  

\begin{figure}
    \centering
    \includegraphics[width=0.99\linewidth]{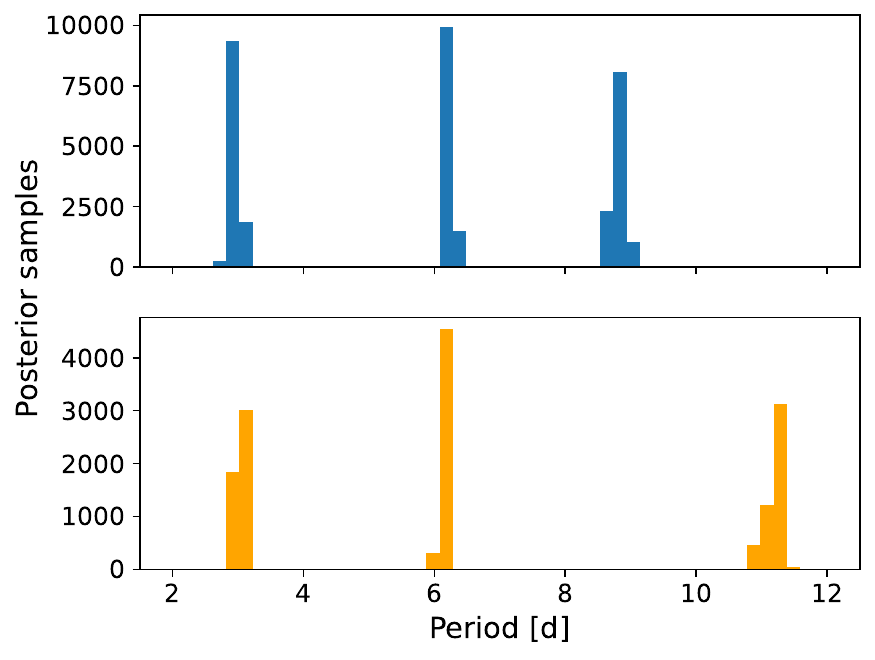}
    \caption{Period posterior histograms for samples with three detected Keplerians (+ KO), with the restriction that periods are $2.5<P_{\rm b}<3.5$, $5.5<P_{\rm c}<6.5$, and \textbf{Top:} $8.5<P_{\rm d}<9.5$, or \textbf{Bottom:} $11<P_{\rm d}<12$. Note that the y-axes use independent scaling. The choice of $P_{\rm d}$ slightly alters the derived period distributions for the two other interior planets.}
    \label{fig:911phist}
\end{figure}

Using a Beta prior in our \textsc{kima} runs identified signals on periods of $\sim3$ and $\sim6$ days, in broad agreement with the results in Sections~\ref{sec:recper} and~\ref{sec:HD28471-rvsol}. However, the third inner planet was seen to either lie on a $\sim9$ or $\sim11$\,d orbit. These two periods were similarly probable, and the two solution `families' are plotted in Figure~\ref{fig:911phist}. This uncertainty also causes splitting of the shortest period: the samples with planet $P_{\rm d}\sim11$\,d prefer higher $P_{\rm b}$, and lower $P_{\rm c}$ than those with $P_{\rm d}\sim9$\,d, evident from the histograms. 

\begin{figure}
    \centering
    \includegraphics[width=0.8\linewidth]{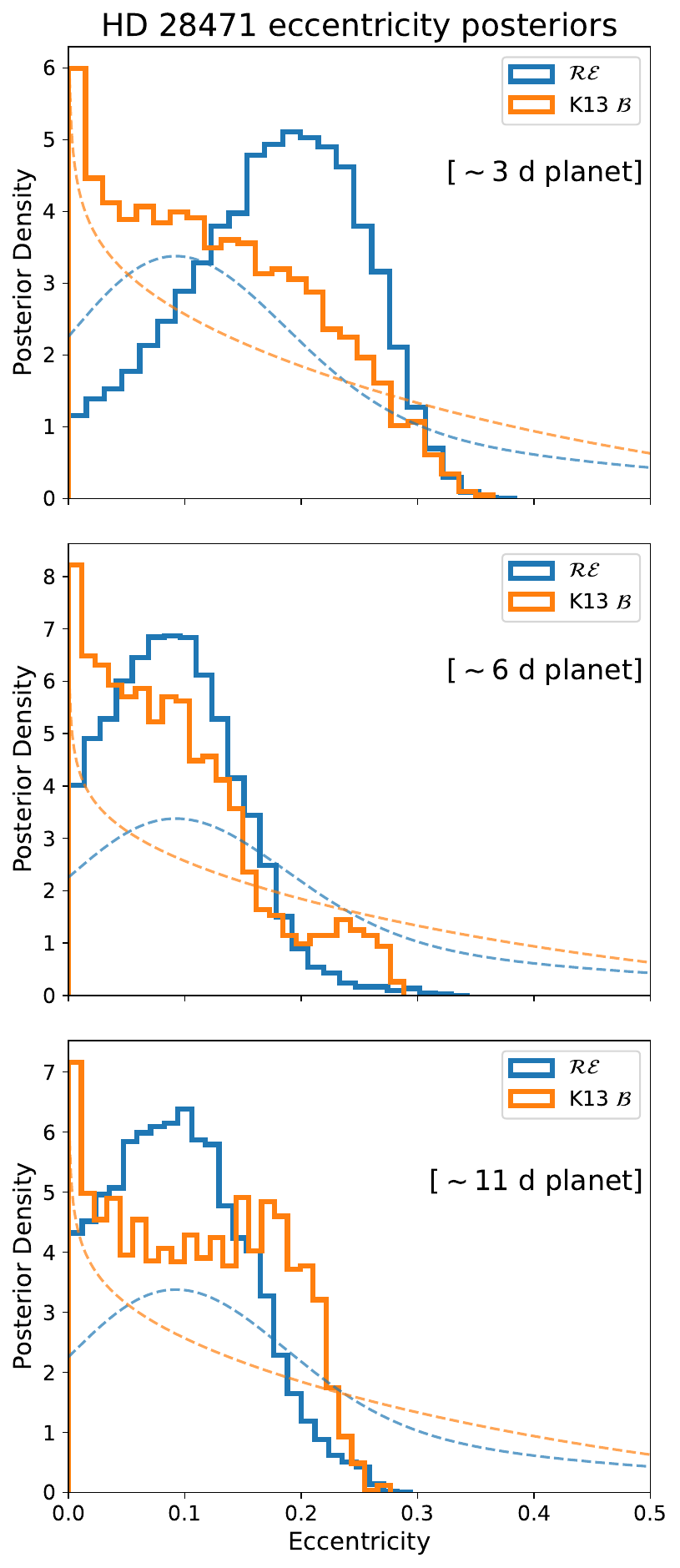}
    \caption{HD\,28471 inner-planet posteriors when using either a Beta eccentricity prior \citep{kipping2013}, or a $\mathcal{RE}$ eccentricity prior \citep{Stevenson-ecc}. The priors clearly have an impact (plotted in dashed lines). The $\sim3$~d planet seems to deviate from the prior, to larger eccentricity values. It may truly be more eccentric, but could also be accounting for some RV variation caused by an additional interior undetected planet.}
    \label{fig:HD28471-compecc}
\end{figure}

To compare the posterior response when using either prior, we have chosen to use only solution families with $P_{\rm d}\sim11$~d, to make sure we are discussing the same candidate planets.
The eccentricity posteriors are shown in Figure~\ref{fig:HD28471-compecc}. The priors are also plotted on these histograms in dashed lines of the same colours. 

The results are different for the two runs, though share some common features. The posteriors are truncated at the same upper eccentricities for either Beta or $\mathcal{RE}$ prior, due to the AMD stability criterion. Eccentricity posteriors for the planets on $\sim6$ and $\sim11$~d orbits seem to reproduce the input eccentricity prior, with slight divergence due to the impact of the AMD condition. Both posteriors for the $\sim3$\,d planet are more distinct from the input $e$ prior. The innermost planet may truly be slightly more eccentric, or it could be that an additional, low-mass undetected planet resides even closer to the star that is not yet formally detected -- and slight RV variation is being absorbed into another planet (see Section~\ref{sec:HD28471-4inner}). In the posteriors for $N_{\rm p}=4$ (+ KO) solutions, the planet orbits are all consistent with circular, even with a $\mathcal{RE}$ prior. The elevated eccentricity of the $\sim3$\,d planet has been removed by adding an interior planet, mopping up some of the RV modulation. More data is required to differentiate between an eccentric planet or two planets on circular orbits \citep[e.g][]{Anglada2010}.

The main take-away from this comparison (other than the fact that using the $\mathcal{RE}$ prior has eliminated a fair amount of period ambiguity) is that on the whole, the posteriors for low-amplitude signals in compact systems can be highly dependent on the prior used. Therefore, one should give this proper consideration and perform thorough experimentation, rather than only using a single prior.

\section{Bisectors, Central line moments, and Bisector offset}\label{app:bisector-offset}

\subsection{HARPS Bisector Offset}

\citet{Trifonov2020} highlighted the need to treat activity indicator data sets as independent data sets due to the fibre exchange, including the BIS/bisector span. In DMPP datasets \citep{Haswell2020}, there is evidence that this is required by comparison of data before and after 2015. This prompted our study of the bisector line profiles themselves, rather than just the BIS value, to see if a shape change confirmed the need for an offset -- and to investigate whether or not any offset is required due to the 2020 re-calibration. Little has been said throughout literature on how the actual bisector profiles will change as a result of these interventions.

The instrumental profile changed significantly during the fibre upgrade, and this also affects the bisectors. Inspecting the bisectors for HD\,28471 (Fig.~\ref{fig:HD28471-bisectors}), there is a noticeable difference in the bisector shape before and after the HARPS fibre upgrade. To investigate if the same effect is seen in other stars, we have studied the bisector profiles in a small number of HARPS archival datasets. 

The change in instrumental profile will be most readily observed for quiet stars, with a low level of chromospheric emission. Astrophysical processes should have a much lower impact on the bisector shapes
for these stars, allowing identification of purely instrumental effects. We began by searching in the DMPP data-set for any stars with sufficient measurements before and after the upgrade. By design, the DMPP sample has sub-basal chromospheric emission, so is a good place to begin \citep[][]{Haswell2020}. The only suitable candidate here was DMPP-TOI-17, which is as of yet unpublished. This is a G3V star \citep{1975mcts.book.....Houk}, suitable for a fairly direct comparison with HD\,28471 (also G-type). To find other suitable quiet candidate stars, we used the catalogue of cool star chromospheric activity created by \citet{BoroSaikia2018}. From the least active G-type stars, two more stars were identified with sufficient temporal coverage to compare with the bisectors of HD\,28471: HD\,125184 (G5IV) and HD\,221420 (G2IV-V). 

\begin{figure*}
    \includegraphics[width=0.33\linewidth]{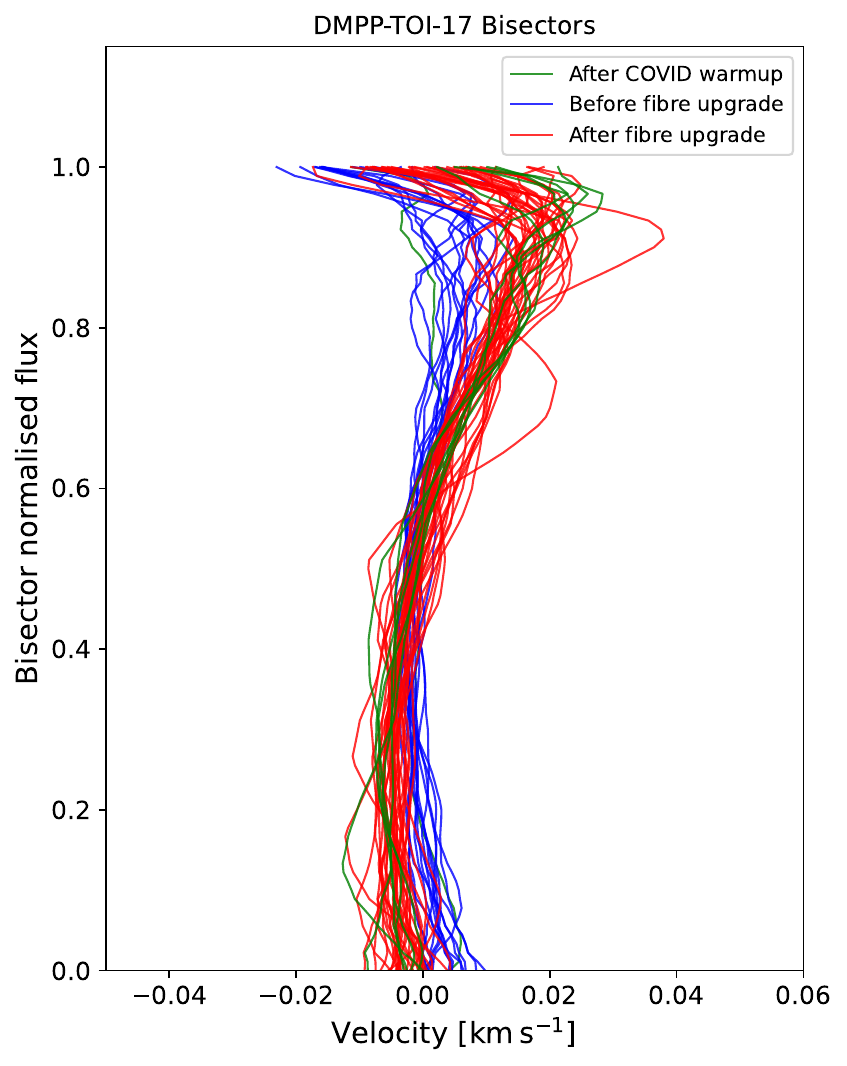}\includegraphics[width=0.33\linewidth]{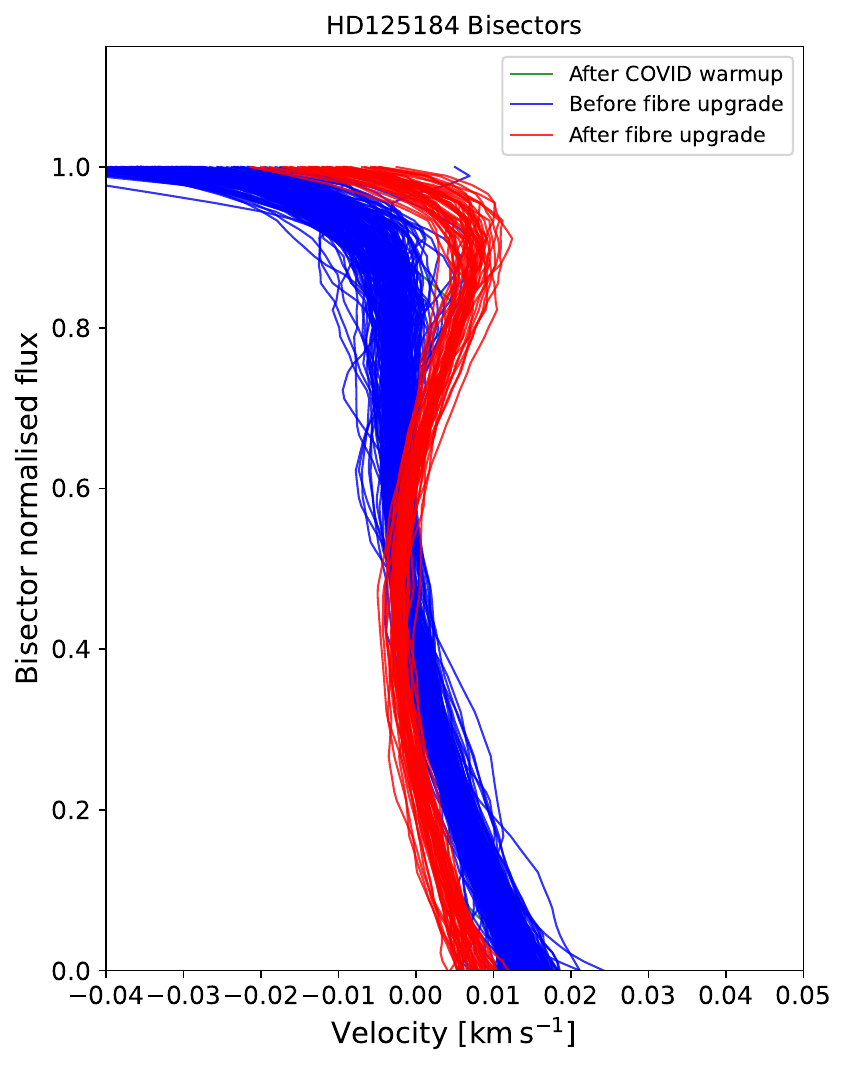}\includegraphics[width=0.33\linewidth]{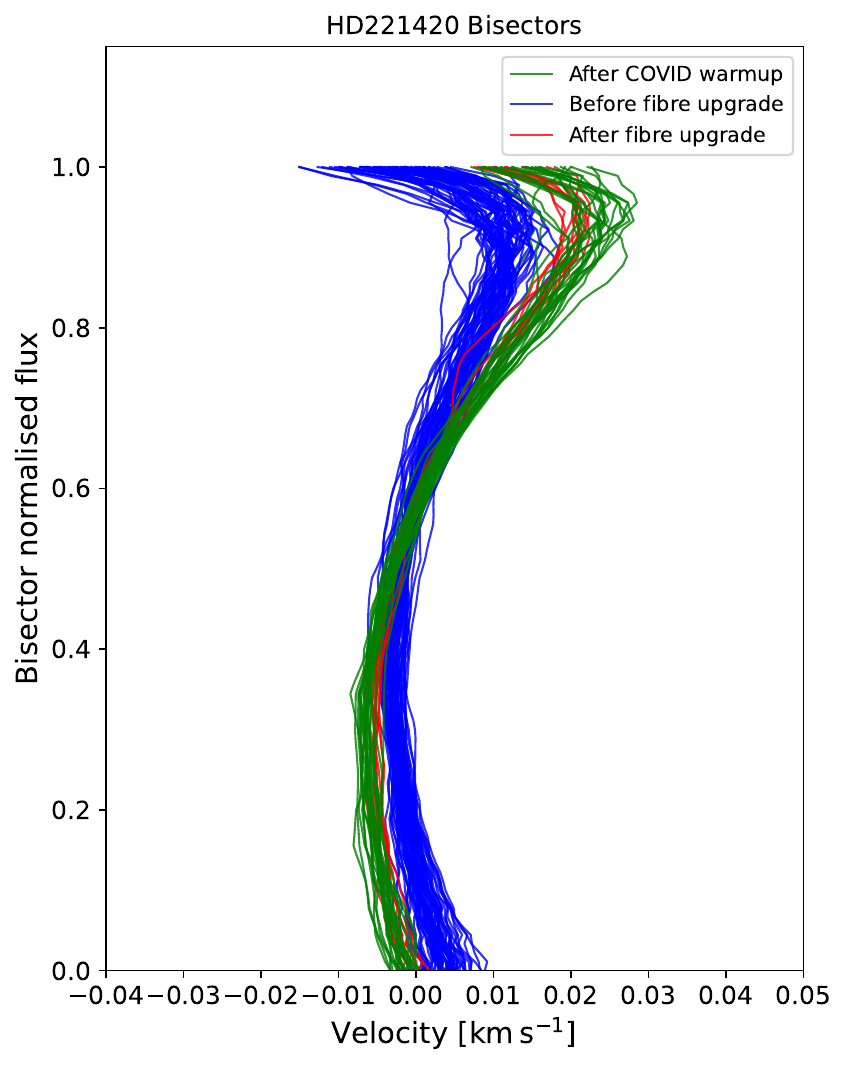}
    \includegraphics[width=0.33\linewidth]{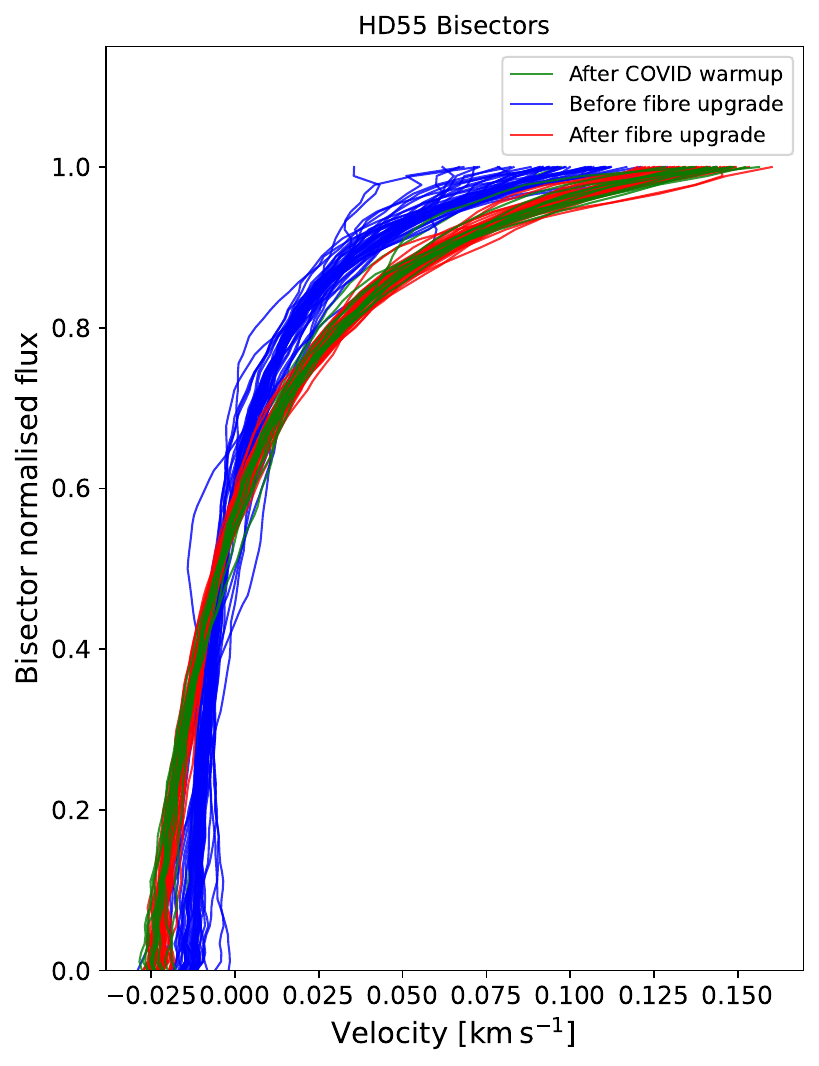}\includegraphics[width=0.33\linewidth]{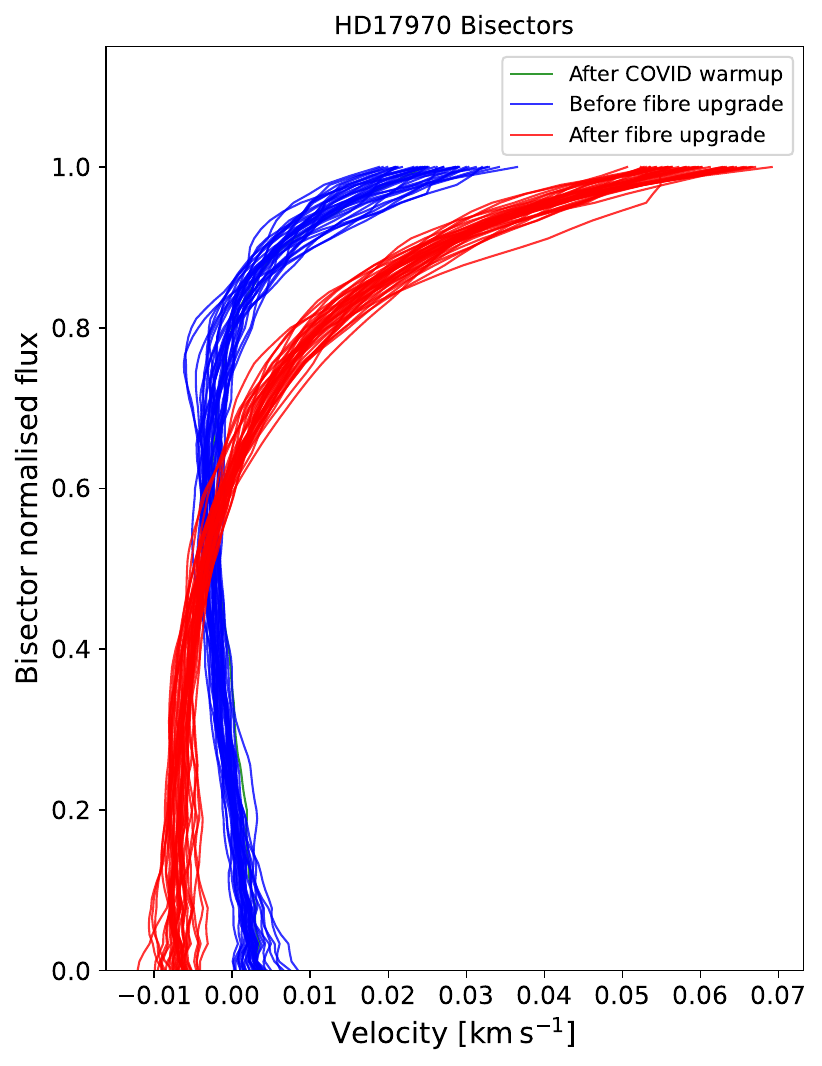}\includegraphics[width=0.33\linewidth]{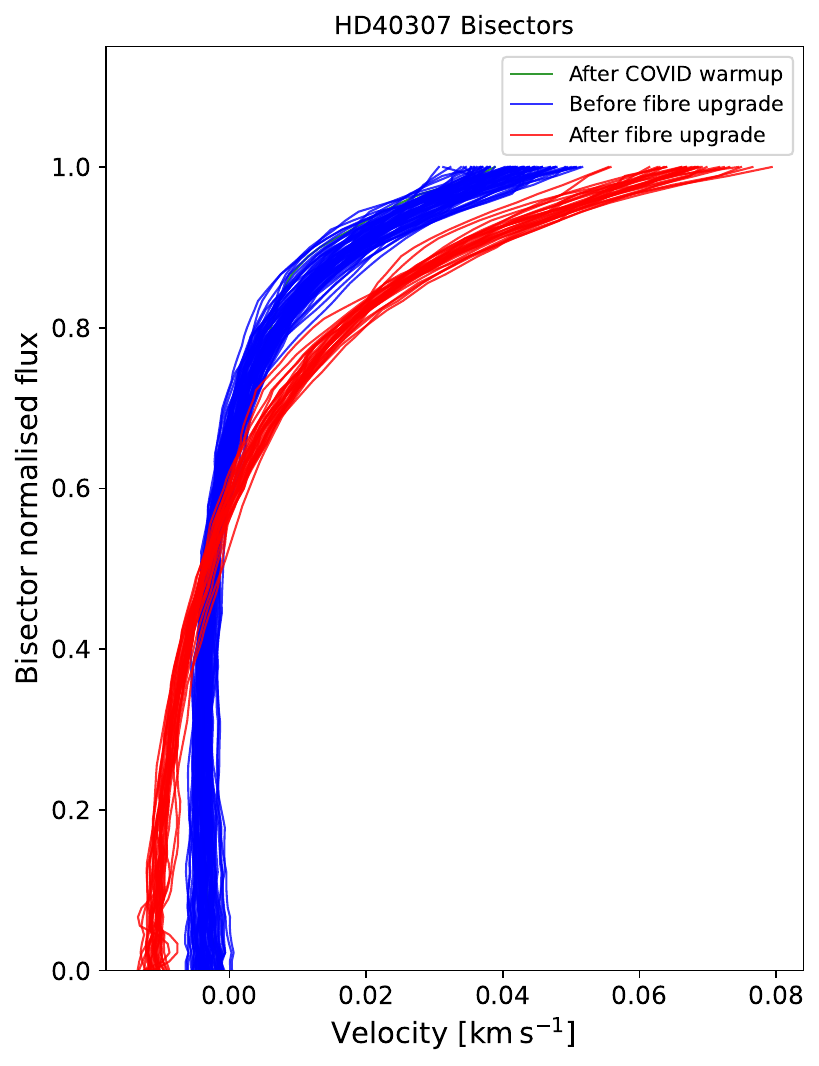}
    \caption{\textbf{Top row:} Bisectors for G-type stars DMPP-TOI-17 (Left), HD125184 (Center), and HD221420 (Right). \textbf{Bottom row:} Bisectors for K-type stars HD\,55 (Left), HD\,17970 (Centre), and HD\,40307 (Right). These plots demonstrate the change in profile caused by the HARPS fibre upgrade in 2015. All lines are subtracted by their corrected CCF RV to highlight the shape change.}
    \label{fig:bisector_comparison}
\end{figure*}

The bisectors for these three stars are accessed from the ESO archive\footnote{Bisector files obtained from the ESO archive, specifying HARPS data only: \href{http://archive.eso.org/wdb/wdb/adp/phase3_spectral/form?phase3_collection=HARPS}{archive.eso.org}}, and plotted in the top row of Fig.~\ref{fig:bisector_comparison}. The number of spectra taken for each ranges from 49 (16+33) for DMPP-TOI-17, to 228 (181+47) for HD125184. Despite a relatively small number of measurements for DMPP-TOI-17, there is an obvious shape difference between pre- and post-upgrade datasets, similar to that seen in Fig.~\ref{fig:HD28471-bisectors}. This is then affirmed by the two other stars. 
The bisector of the mean CCF will change as a result of the illumination and line spread function being different when using either circular or octagonal fibres. This justifies treating bisector span measurements as two independent datasets, as noted by \citet{Trifonov2020}.

Measurements taken after the COVID warm-up are of a very similar shape to those taken just before, indicating that there may be no need for an offset between these two data sets. Indeed, analysis of any pre-correction BIS data reveals that the post-COVID set has a very similar median to those recorded before March 2020.

To confirm this is not a feature solely for G-type stars, we have selected three K-type stars in the \citet{BoroSaikia2018} catalogue with low chromospheric activity, that have amassed a substantial record of HARPS spectra before and after the fibre upgrade. These are HD\,55 (K4.5V), HD\,17970 (K2V), and HD\,40307 (K2.5V). These K-stars have bisectors with a different shape to the G-stars, due to the convolution mask used in the CCF technique employed by the HARPS \textsc{DRS} (G2 vs. K5). The profile changes due to fibre upgrade are even more pronounced for the K-stars studied, with less overlap between the two regimes. As before, the post-COVID observations do not show a similar shape change (Fig~\ref{fig:bisector_comparison}, bottom left panel).

\subsection{Line Profile Behaviour in HD 28471}

As there are three effective data-sets for HD\,28471, offsets between these sets may be present. Normalising the BIS measurements \citep{Queloz2001} by subtracting off each set's median removes most of the opportunity to examine long-period activity periodicities effectively. The individual sets have baselines of 3054, 763, and 9~d respectively. The longest baseline however corresponds to only 23 sparsely sampled measurements. Analysing the BIS time-series, we find no periodicities corresponding to the long-period planet signal. On short timescales, where period identification should be more successful, there are no identified peaks in the GLS periodogram (Fig.~\ref{fig:HD28471-DRSindicator-GLS}). All peaks are detected far below the $10\%$ FAP level. Correlations between the BIS and the measured RV are insignificant, shown in Figure~\ref{fig:HD28471-DRSindicator-correlations}. This supports the interpretation that HD\,28471 is truly host to a multi-planet system. 

Similarly, central line moments (CLMs) do not provide any firm evidence that the RV signatures observed in HD\,28471 are caused by stellar activity. CLMs use a larger amount of CCF information than the BIS, with the third moment (skewness of the line profile) often providing a stronger correlation with RV than the BIS for activity induced variations \citep[for more information on CLMs, see][]{Barnes-moments}. 

Observations taken after the COVID warm-up display bisectors with a similar shape to those taken just before, indicating that perhaps there is no need for an offset between these two data-sets. Indeed, analysis of the original BIS time series reveals that the post-COVID set has a very similar median to those taken pre-COVID. The change in instrumental line profile caused by the substantial intervention during the fibre upgrade has affected the bisectors -- though the same may not be true for the recalibration after the COVID-19 HARPS shutdown.

The lack of BIS evidence for an activity interpretation reflects what is seen upon investigation of the bisectors themselves. Taking the last two data-sets together as one set, and re-performing period analysis, we see similar results: no periodicities are detected with significant power. The existence of a COVID offset in the bisectors is therefore of no importance for this particular case, though offsets of this type could require attention for other activity indicators.

\section{dLW correction}\label{app:dLW-trend}

It is already established that measurements of the FWHM require a polynomial correction in the pre-upgrade era, due to the slowly drifting focus of the HARPS instrument \citep{2018A&A...Dumusque,2021MNRAS...Costes,Stevenson2023-DMPP3}. The differential line width was created by \citet{SERVALpaper} to provide an analogous measurement to FWHM, for non-CCF based data reduction processes. It would therefore be expected that a drift should also be observed in dLW. HD\,28471 dLW data does appear to also show a trend in the pre-upgrade data. To investigate this trend, we have derived a correction for pre-upgrade dLW, in a similar manner to the FWHM correction from \citet{2021MNRAS...Costes}. We have chosen to use a similar sample of stars as \citet{2021MNRAS...Costes}, retaining those with a tractable number of observations for storage space, download speeds, and \textsc{serval} reduction time (nominally $\lesssim1000$ spectra).  We also use stars with spectra acquired for the investigation into bisector offsets above (Appendix~\ref{app:bisector-offset}), as these purposefully have plentiful observations in both the pre- and post-fibre-upgrade eras. The samples of G- and K-type stars used in this section are listed in Tables~\ref{tab:dLW_correction_stars_G} and~\ref{tab:dLW_correction_stars_K}, respectively.

\renewcommand{\arraystretch}{1.4}
\begin{table}
    \centering
    \begin{tabular}{lcc}
    \hline
        Star & SpT & Number of dLW measurements \\
        \hline
         HD\,1461 & G3V & 477 \\
         HD\,10180 & G1V & 336 \\
         HD\,20003 & G8V & 203 \\
         HD\,20807 & G1V & 366 \\
         HD\,38858 & G2V & 252 \\
         HD\,45364 & G8V & 343 \\
         HD\,59468 & G6.5V & 527 \\
         HD\,69830 & G8V & 729 \\
         HD\,73524 & G0V & 214 \\
         HD\,90156 & G5V &  160 \\
         HD\,96700 & G1V & 355 \\
         HD\,106116 & G5V & 141 \\
         HD\,125184 & G5/6V & 223 \\
         HD\,136352 & G3/5V & 477 \\
         HD\,157172 & G8.5V & 128 \\
         HD\,161098 & G8V & 320 \\
         HD\,189567 & G2V & 651 \\
         HD\,221420 & G2IV-V & 97 \\
         \hline
         
    \end{tabular}
    \caption{G-type stars used for correction of differential line widths. Information from \citet{2021MNRAS...Costes} and SIMBAD.}
    \label{tab:dLW_correction_stars_G}
\end{table}

\renewcommand{\arraystretch}{1.4}
\begin{table}
    \centering
    \begin{tabular}{lcc}
    \hline
        Star & SpT & Number of dLW measurements \\
        \hline
         HD\,55 & K4.5V & 98 \\
         HD\,13060 & K1V & 143 \\
         HD\,13808 & K2V & 373 \\
         HD\,17970 & K2V & 74 \\
         HD\,26965 & K0V & 968 \\
         HD\,39194 & K0V & 365 \\
         HD\,65277 & K4V & 80 \\
         HD\,72673 & K1V & 895 \\
         HD\,82516 & K2V & 199 \\
         HD\,85390 & K1.5V & 172 \\
         HD\,101930 & K2V & 108 \\
         HD\,104067 & K2V & 178 \\
         HD\,109200 & K1V & 1038 \\
         HD\,113538 & K9V & 131 \\
         HD\,125595 & K4V &  273\\
         HD\,136713 & K2V & 111 \\
         HD\,144628 & K2V & 518 \\
         HD\,154577 & K2.5V & 1000 \\
         HD\,204941 & K2V & 126 \\
         HD\,215152 & K3V & 518 \\
         \hline
         
    \end{tabular}
    \caption{K-type stars used for correction of differential line widths. Information from \citet{2021MNRAS...Costes} and SIMBAD.}
    \label{tab:dLW_correction_stars_K}
\end{table}

\begin{figure}
    \centering
    \includegraphics[width=0.99\linewidth]{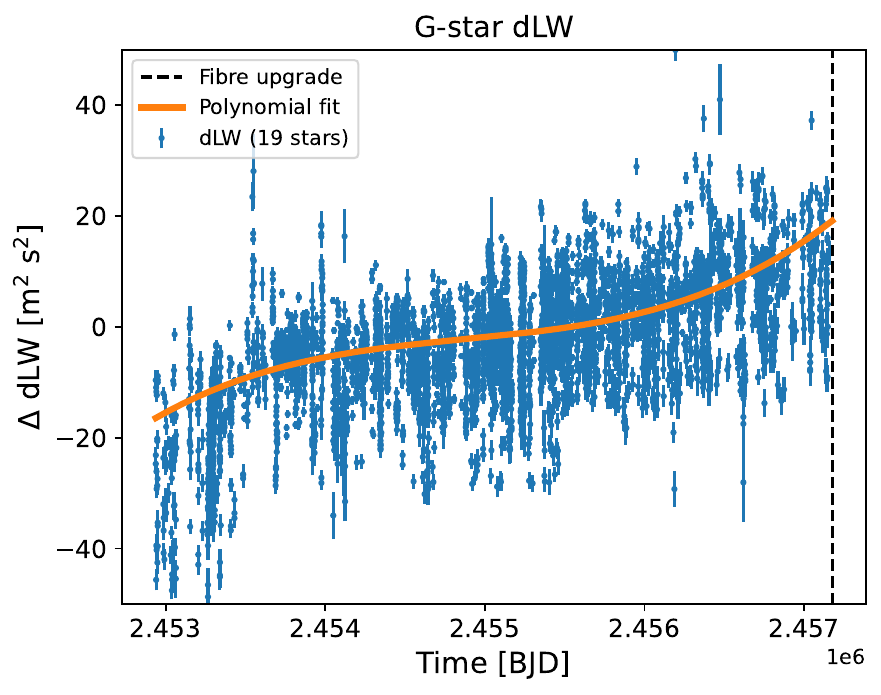}
    \caption{The mean-subtracted dLW computed for 19 G-type stars, with the polynomial best-fit drawn in orange. The equivalent plot for K-type stars shows similar behaviour. The fibre upgrade date is indicated by a black vertical dashed line.}
    \label{fig:dlWcorrection}
\end{figure}

\begin{figure}
    \centering
    \includegraphics[width=0.90\linewidth]{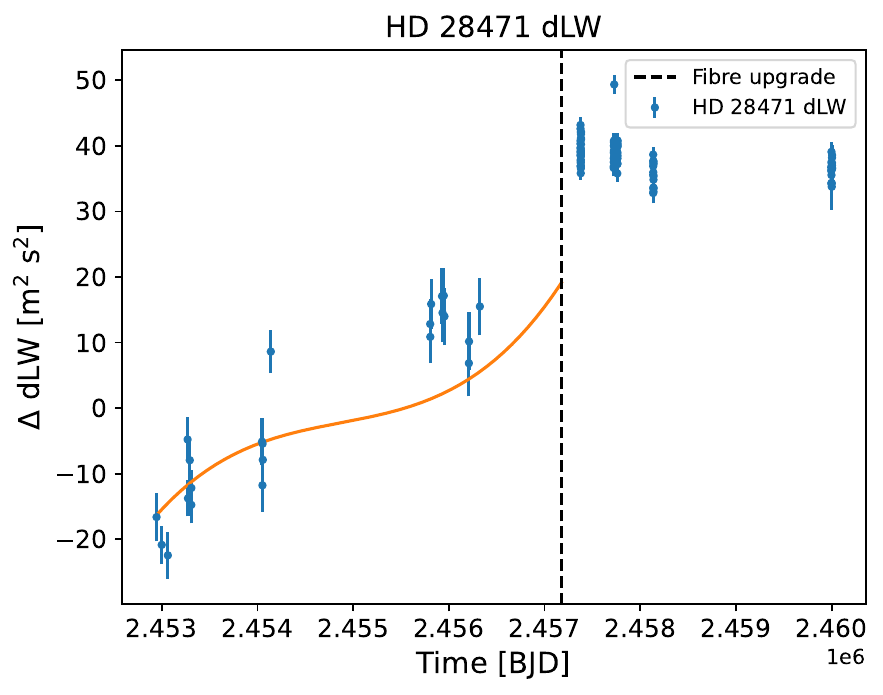}
    \includegraphics[width=0.90\linewidth]{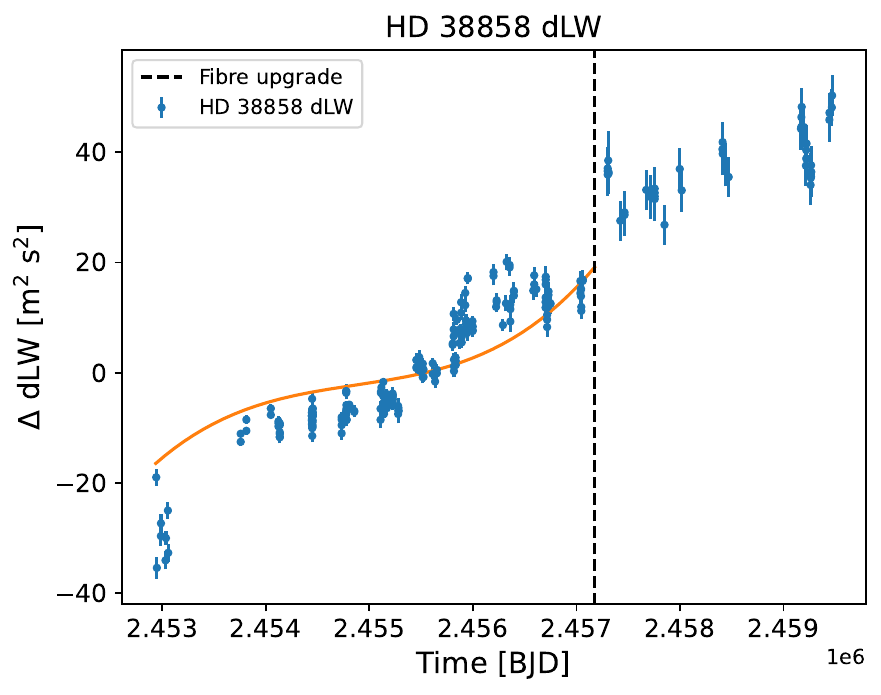}
    \includegraphics[width=0.90\linewidth]{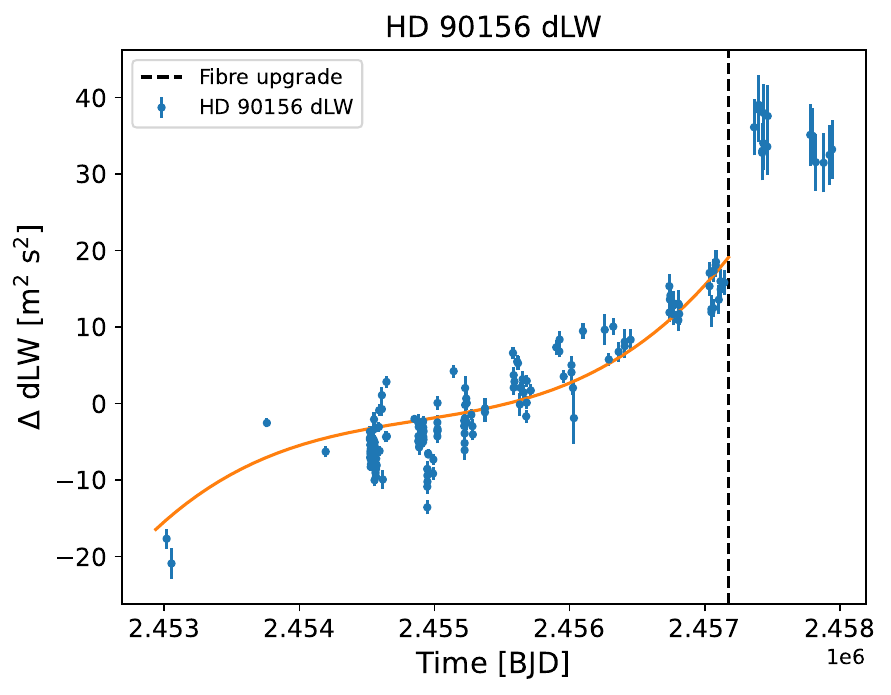}
    \caption{Differential line width time-series for a small sample of G-type stars. HD\,28471 is shown alongside two other stars to assess the individual trends. At the earliest epochs, there seems to be a step-change in the dLW, informing the use of a third-order polynomial to correct for the drift.}
    \label{fig:dLWcorrexamples}
\end{figure}

In their correction of FWHM drift, \citet{2021MNRAS...Costes} combined all measurements (mean subtracted for each star) and fitted a second-order polynomial to the data (with polynomial order chosen for simplicity). We employed a similar approach here, but also assessed the fits when including other polynomial orders -- eventually electing to use a third-order polynomial. This choice was informed by inspection of both the combined dataset, and the observations for each individual star. There seems to be a steeper drop off before BJD$\sim 2453500$, necessitating a third order fit (see Figs.~\ref{fig:dlWcorrection} and~\ref{fig:dLWcorrexamples}). 

The (mean-subtracted) dLW is corrected according to Equation~\ref{eqn:dLW_correction}:
\begin{equation}
    \label{eqn:dLW_correction}
    \textrm{dLW}_{\textrm{corrected}} = \textrm{dLW}_{\textsc{serval}} - (a t^{3} + b t^{2} + c t^{1} + d),
\end{equation}

\noindent where $t$ is the Julian date of observation, and $a,b,c$ and $d$ are coefficients of the fit, listed in Table~\ref{tab:dLW_correction_params}. We have performed a polynomial fit for the samples of G and K-type stars (with their own sets of parameters in Table~\ref{tab:dLW_correction_params}). Repeating the process for a sample of F-type stars would a worthwhile extension to this work. The polynomial is subtracted from pre-upgrade dLW values to correct for the drifting focus of HARPS, alongside the required normalisation between the two datasets.

\begin{table}
    \centering
    \begin{tabular}{lccc}
    \hline
        Coefficient & G-type stars & K-type stars\\
        \hline
        a  & $1.21 (\pm 0.00006)\times 10^{-9}$ & $2.65 (\pm 0.0001)\times 10^{-9}$ \\
        b  & $-8.96 (\pm 0.0004)\times 10^{-3}$ & $-1.95(\pm 0.00007)\times 10^{-2}$ \\
        c  & $2.20 (\pm 0.0001)\times 10^{4}$ & $4.79 (\pm 0.0002)\times 10^{4}$ \\
        d  & $-1.80 (\pm 0.00009)\times 10^{10}$ & $-3.92 (\pm 0.0001)\times 10^{10}$ \\
    \end{tabular}
    \caption{Coefficients used in Equation~\ref{eqn:dLW_correction} to correct pre-upgrade drift in dLW from HARPS observations. Calculated using \textsc{numpy}'s \texttt{polyfit} function, with order $=3$.}
    \label{tab:dLW_correction_params}
\end{table}

\section{Activity Correlation plots}
\begin{figure*}
    \centering
    \includegraphics[width=0.80\linewidth]{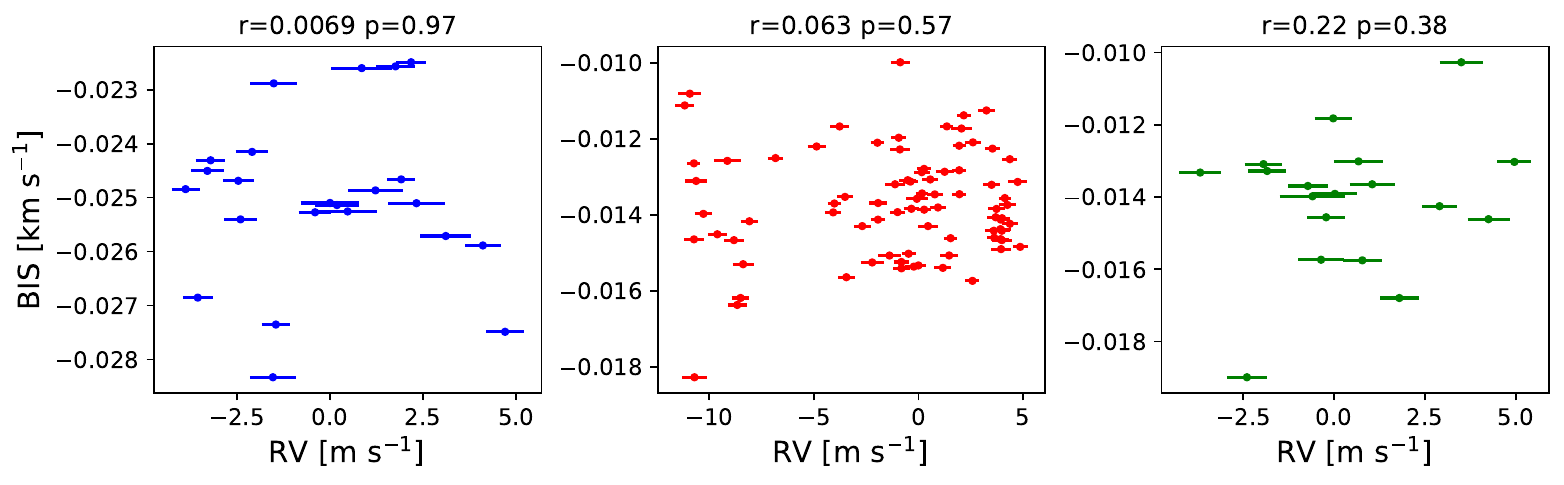}
    \includegraphics[width=0.80\linewidth]{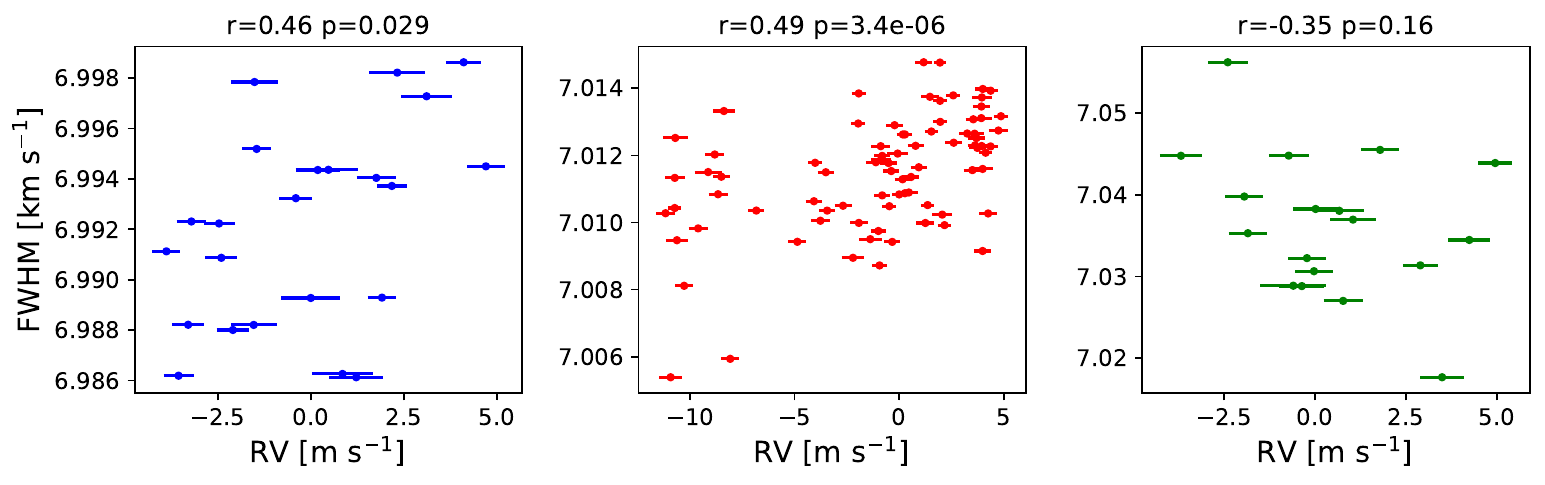}
    \includegraphics[width=0.80\linewidth]{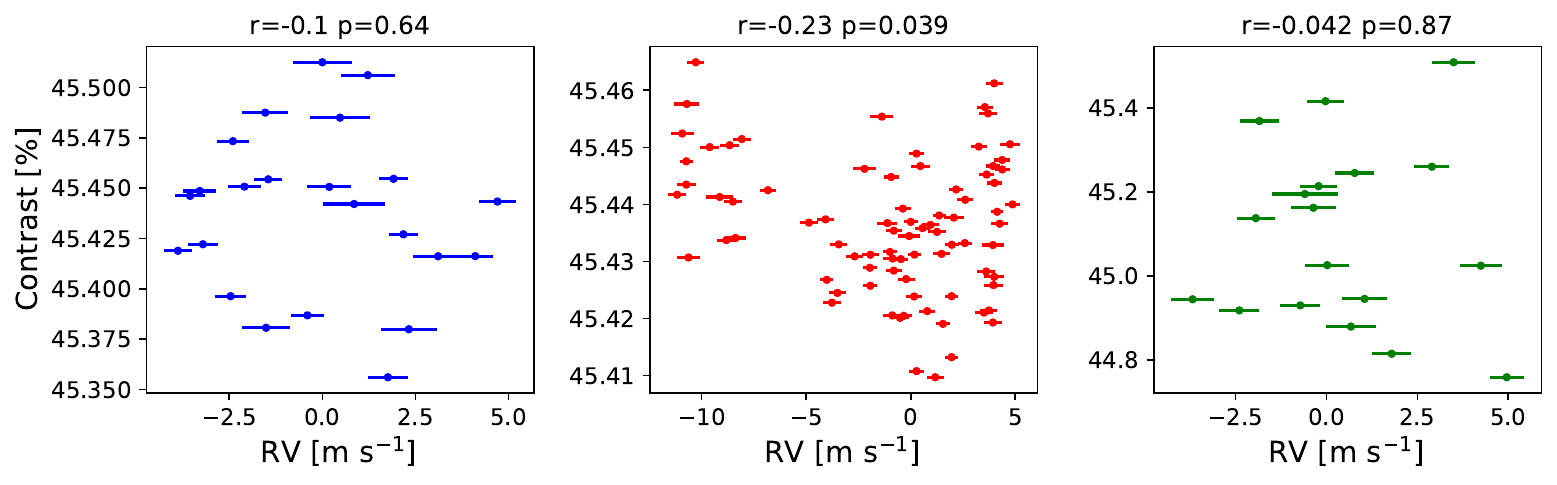}
    \includegraphics[width=0.80\linewidth]{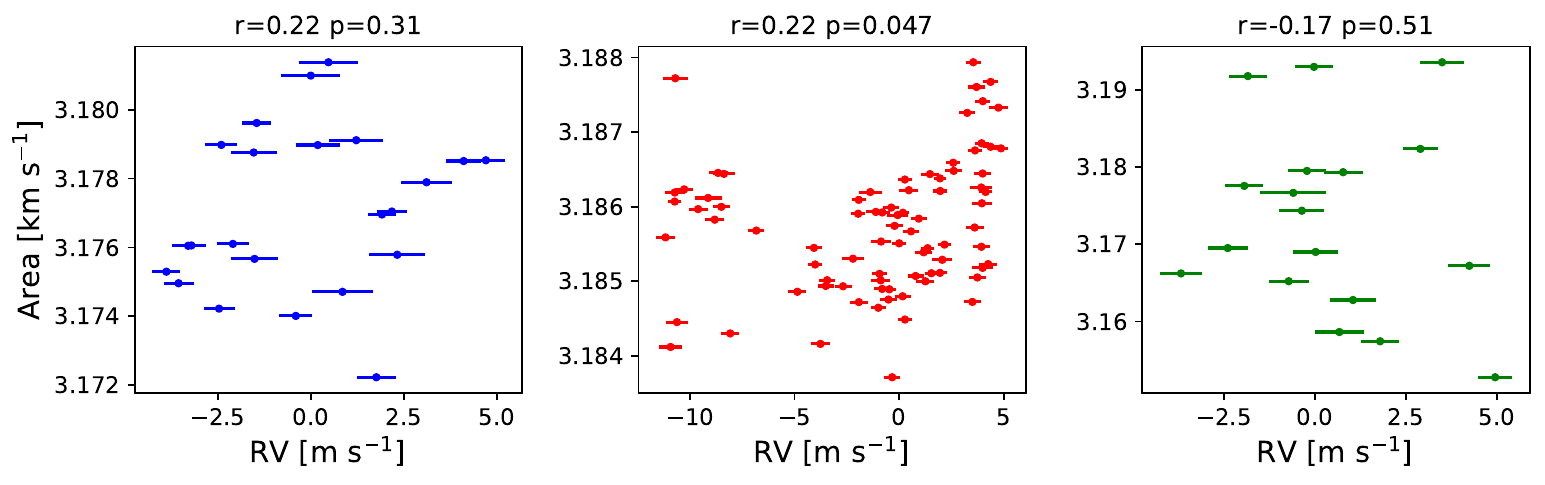}
    \caption{Correlations between RVs and activity indicators derived via the cross-correlation technique employed in the HARPS \textsc{drs}, extracted from the observations by running \textsc{serval}. RV reduction is from \textsc{s-bart}. \textbf{Columns} are split by data-set, to analyse correlations without the possible influence of instrumental offsets. \textbf{Blue:} pre-upgrade, \textbf{red:} post-upgrade, \textbf{green:} post-COVID. Each \textbf{row} shows the three sets for one activity indicator, with BIS, FWHM, Contrast, and Area from \textbf{top to bottom}, respectively.}
    \label{fig:HD28471-DRSindicator-correlations}
\end{figure*}

\begin{figure}
    \centering
    \includegraphics[width=1\linewidth]{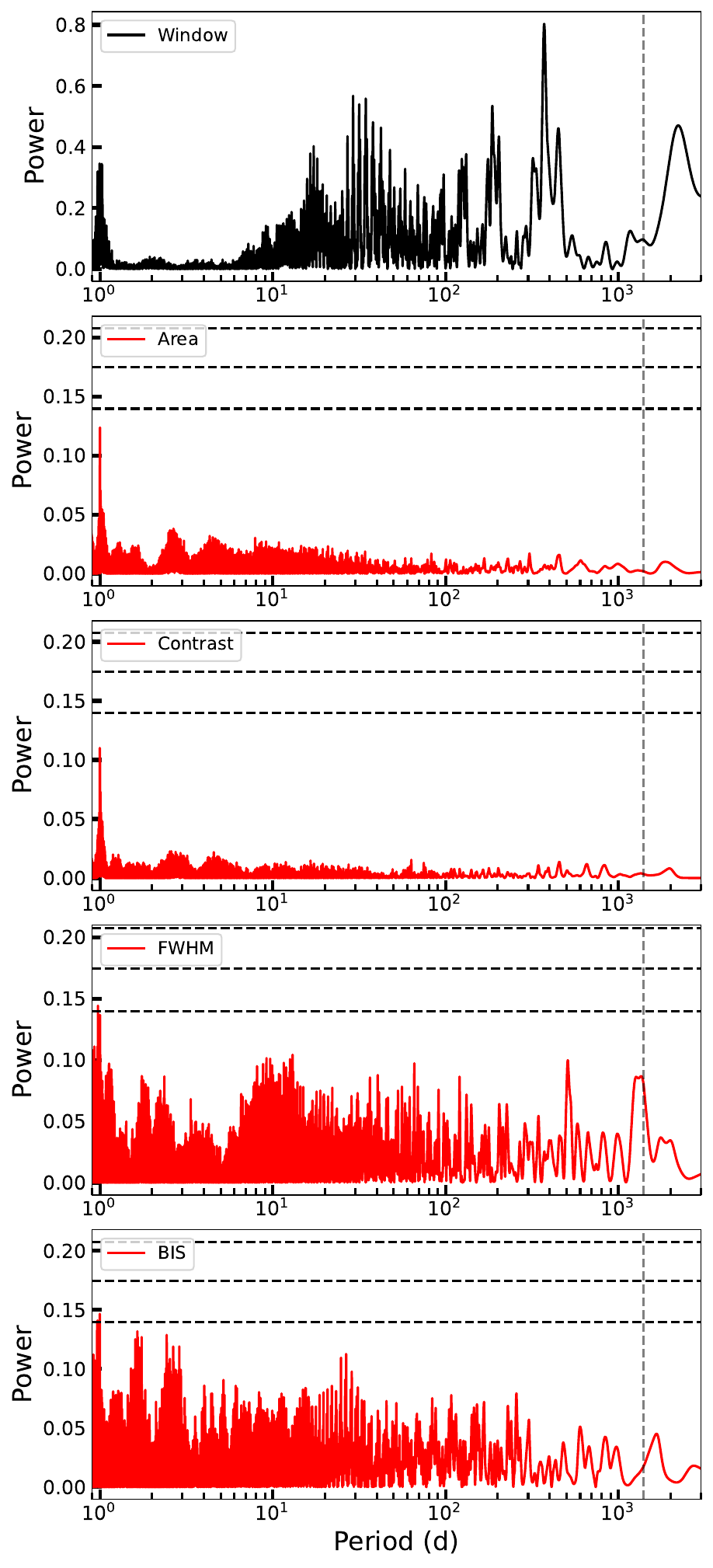}
    \caption{GLS periodograms for \textsc{drs} CCF-based indicators, and the window function (black plot) showing any influence the time sampling has on period recovery \textbf{Top to bottom:} Window function, Area, Contrast, FWHM, BIS. Vertical dashed lines are plotted at the potential period of the long-period signal, to see if this periodicity is detected in activity indicator periodograms. The 10, 1, and 0.1\% false-alarm levels are denoted by horizontal dashed lines.}
    \label{fig:HD28471-DRSindicator-GLS}
\end{figure}

We analysed the CCF Contrast and Area in a similar way to the analysis reported for FWHM in Section~\ref{sub:lineprofiles}. The values were corrected in an analogous manner to FWHM \citep{2021MNRAS...Costes}. No significant periodicities are detected (Fig.~\ref{fig:HD28471-DRSindicator-GLS}), but a correlation with $p<0.05$ is seen in post-upgrade data (Fig.~\ref{fig:HD28471-DRSindicator-correlations}). By combining contrast and FWHM, one can calculate the CCF Area, another activity indicator \citep{2019.Collier.Cameron}. The same analysis was performed on the Area, where an only-just-significant ($p=0.047$) correlation is seen with radial velocities (Fig.~\ref{fig:HD28471-DRSindicator-correlations}) -- mostly due to the correlation observed in FWHM. Interestingly, the area is less correlated with RV than the datasets combined to make it, indicating the RV correlation in FWHM and Contrast may not be changing exactly in phase with each other. Visually, these `significant' $p$-values do not inspire one to place much faith in the actual existence of any correlations.

\begin{figure*}
    \centering
    \includegraphics[width=0.80\linewidth]{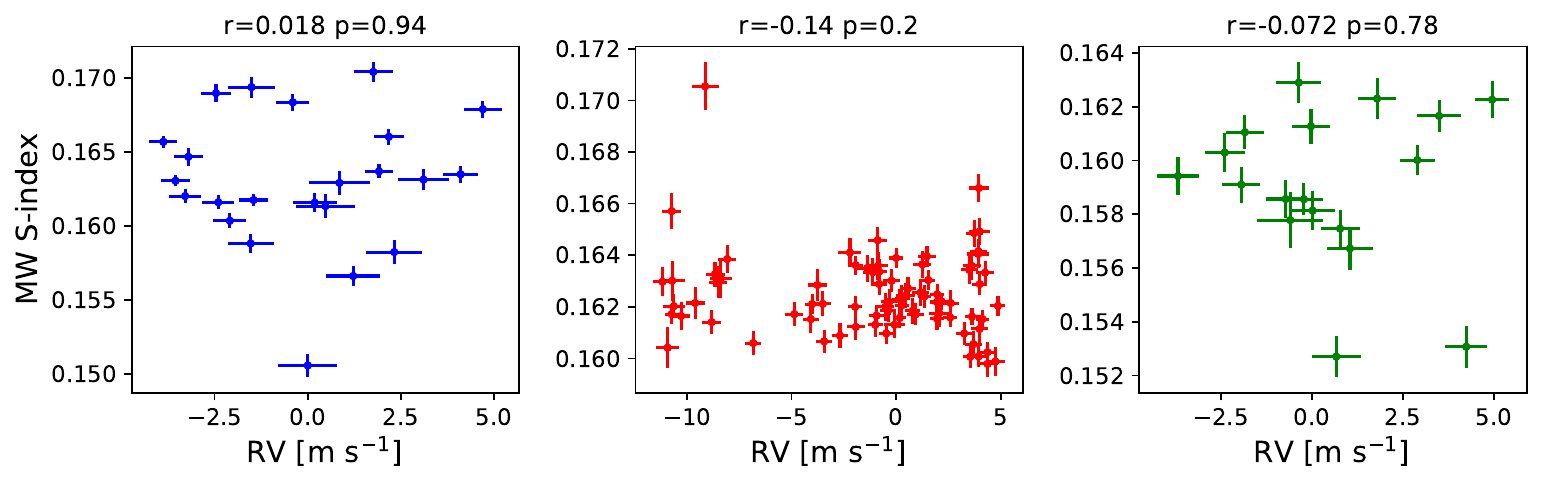}
    \includegraphics[width=0.80\linewidth]{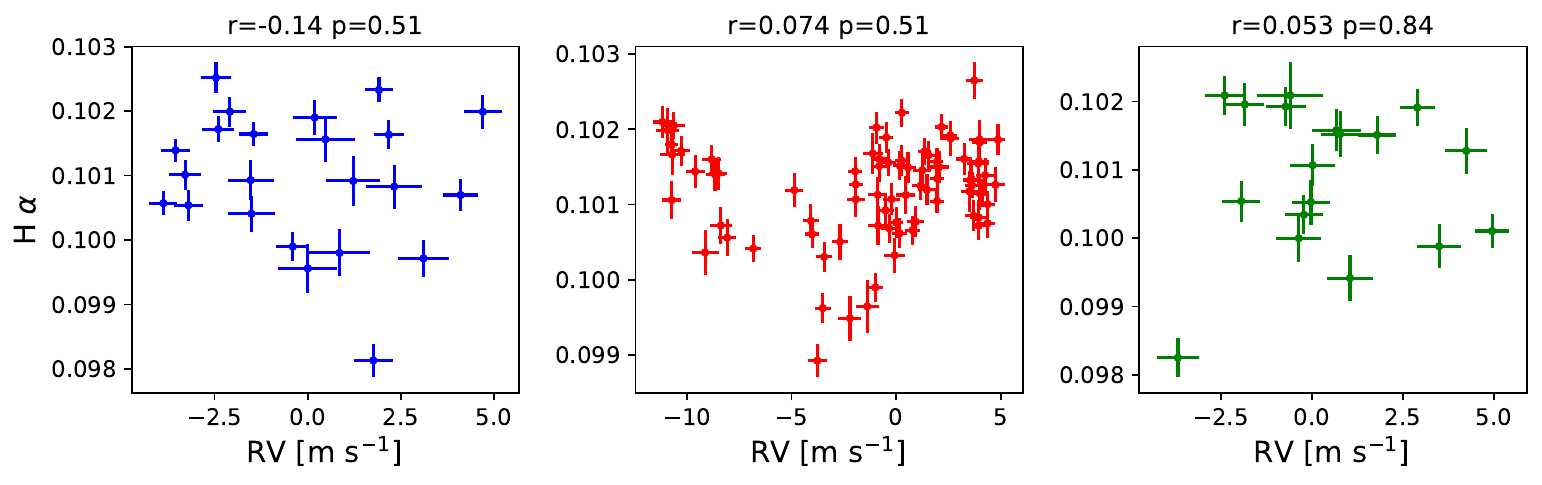}
    \includegraphics[width=0.80\linewidth]{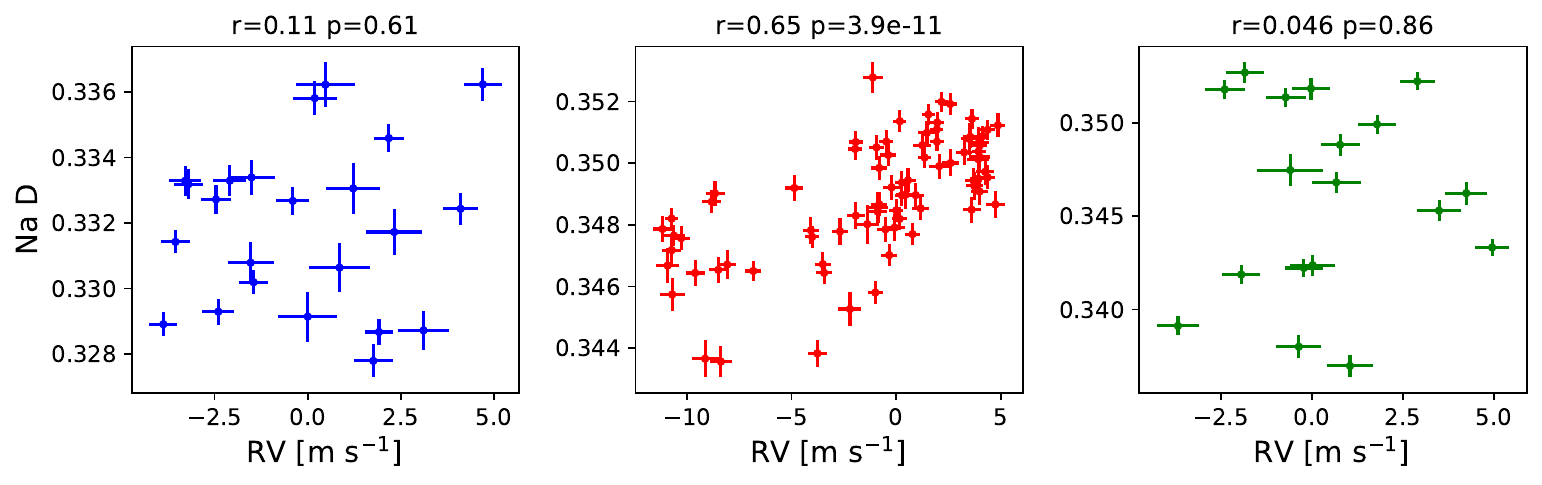}
    \caption{Correlations between RVs and elemental line strength activity indicators derived with the \textsc{actin2} software. RV reduction is from \textsc{s-bart}. \textbf{Columns} are split by data-set, to analyse correlations without the possible influence of instrumental offsets. \textbf{Blue:} pre-upgrade, \textbf{red:} post-upgrade, \textbf{green:} post-COVID. Each \textbf{row} shows the three sets for one activity indicator, with S-index, H$\alpha$, and Na\,D from \textbf{top to bottom}, respectively.}
    \label{fig:HD28471-ACTINindicator-correlations}
\end{figure*}

\begin{figure*}
    \centering
    \includegraphics[width=0.80\linewidth]{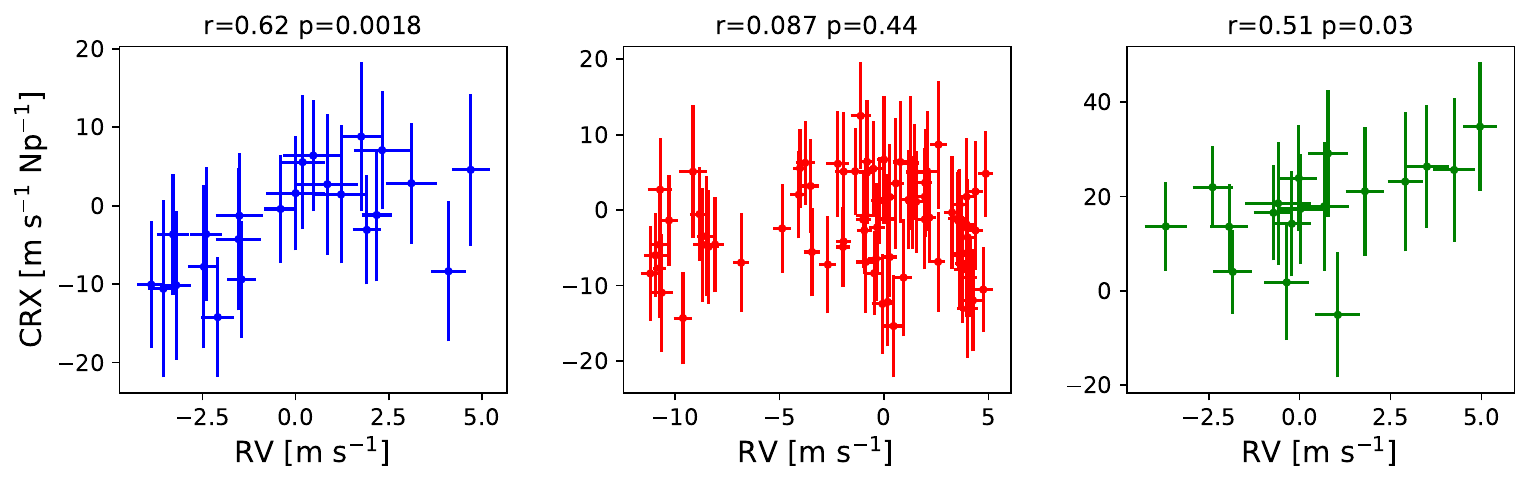}
    \includegraphics[width=0.80\linewidth]{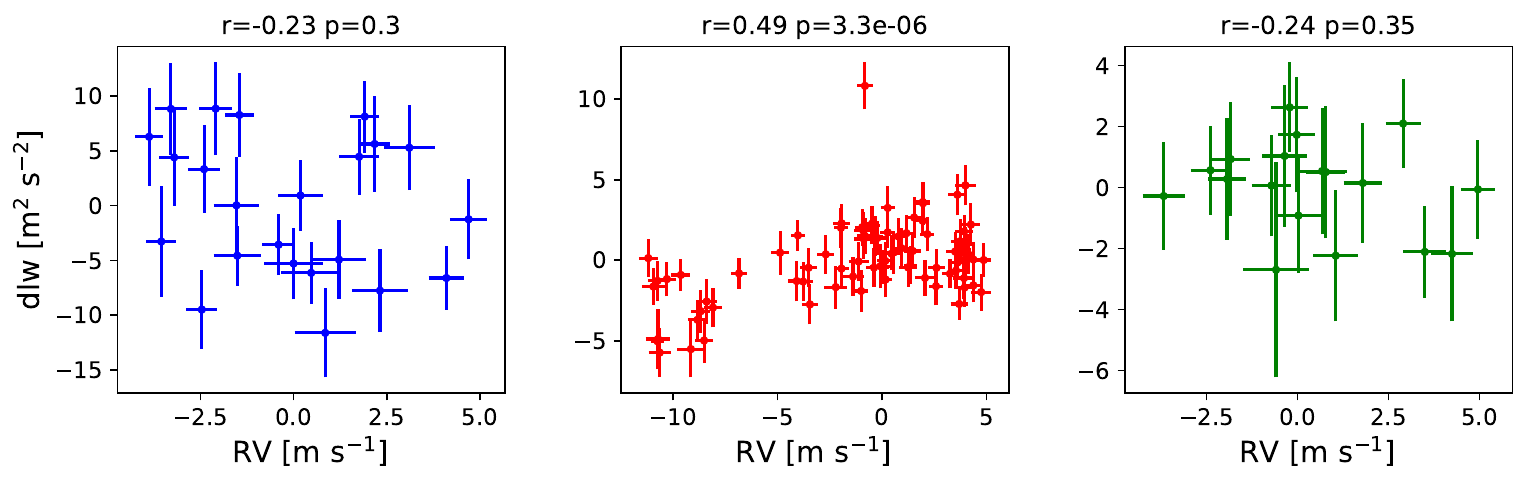}
    \caption{Correlations between RVs and indicators derived with the \textsc{serval} software. RV reduction is from \textsc{s-bart}. \textbf{Columns} are split by data-set, to analyse correlations without the possible influence of instrumental offsets. \textbf{Blue:} pre-upgrade, \textbf{red:} post-upgrade, \textbf{green:} post-COVID. Each \textbf{row} shows the three sets for one activity indicator: CRX (\textbf{top}) and dLW (\textbf{bottom}).}
    \label{fig:HD28471-servalindicator-correlations}
\end{figure*}

\section{Activity indicator periodograms}

\begin{figure}
    \centering
    \includegraphics[width=1\linewidth]{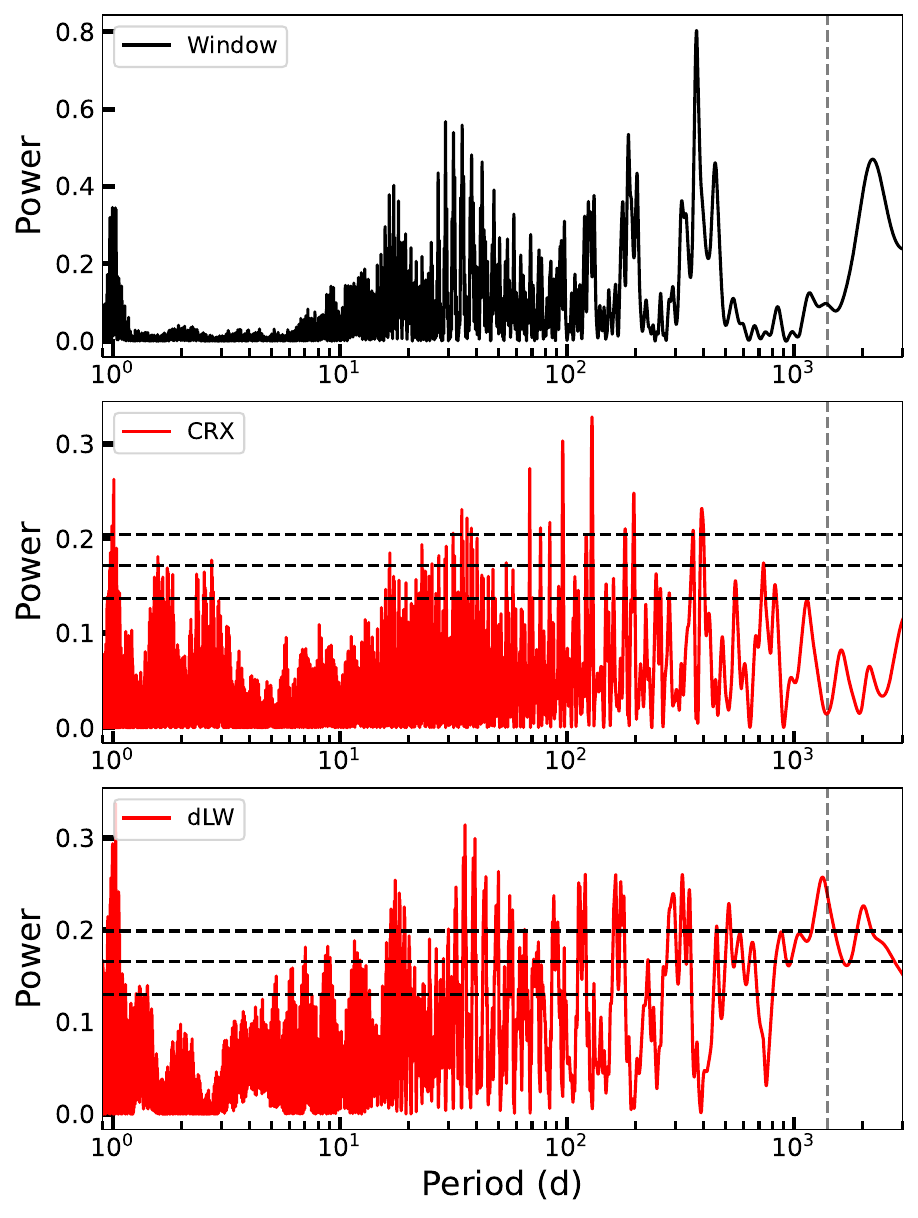}
    \caption{GLS periodograms for \textsc{serval}-derived activity indicators. \textbf{Top:} Window function. \textbf{Middle:} CRX. \textbf{Bottom:} dLW. Vertical dashed lines are plotted at the potential period of the long-period signal, to see if this periodicity is detected in activity indicator periodograms. The 10, 1, and 0.1\% false-alarm levels are denoted by horizontal dashed lines.}
    \label{fig:HD28471-servalindicator-GLS}
\end{figure}

\begin{figure}
    \centering
    \includegraphics[width=1\linewidth]{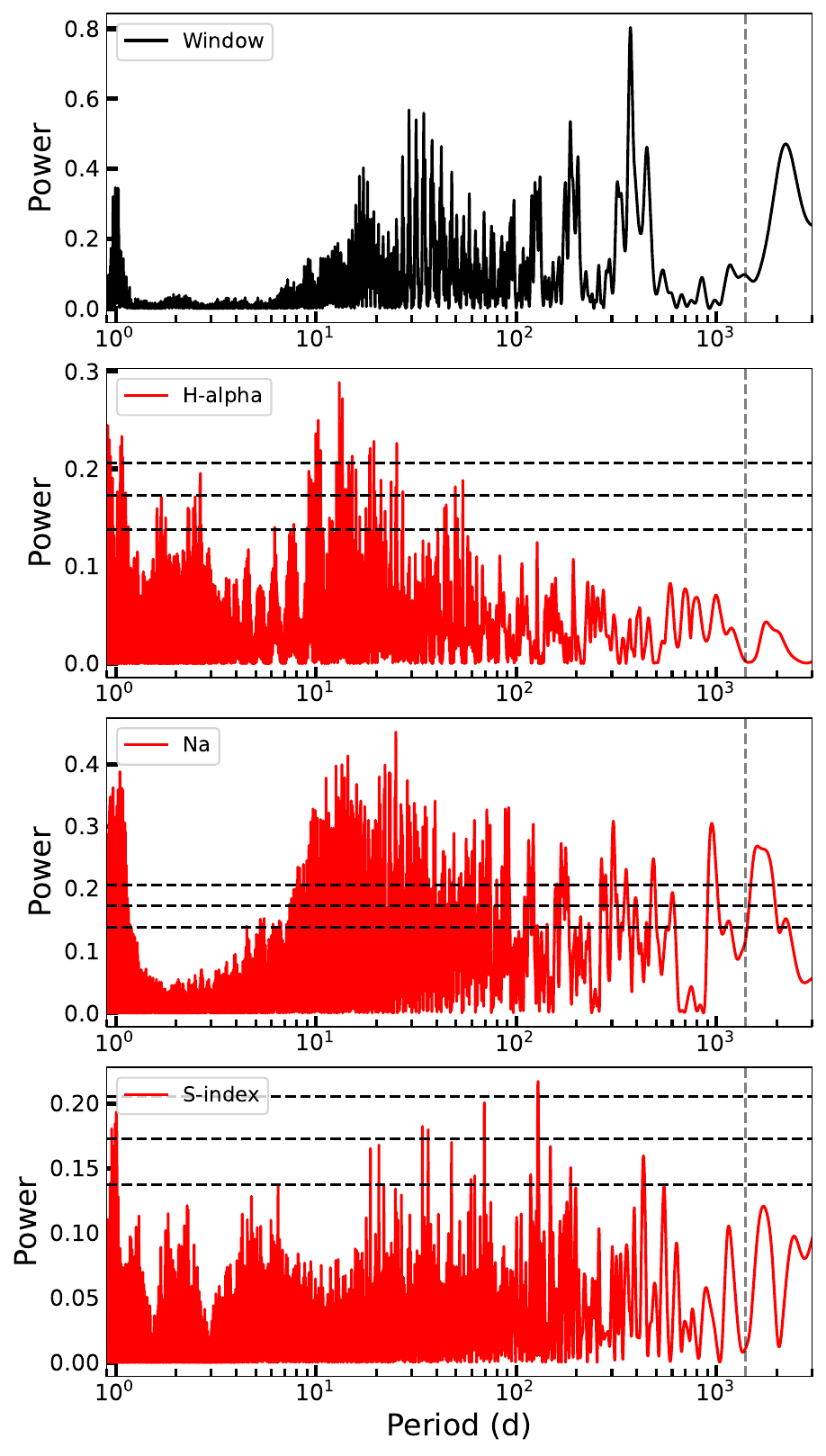}
    \caption{GLS periodograms for \textsc{actin2} elemental line strength indicators. \textbf{Top to bottom:} Window function, H$\alpha$, Na doublet, and MW S-index. Vertical dashed lines are plotted at the potential period of the long-period signal, to see if this periodicity is detected in activity indicator periodograms. The 10, 1, and 0.1\% false-alarm levels are denoted by horizontal dashed lines.}
    \label{fig:HD28471-ACTINindicator-GLS}
\end{figure}

%%%%%%%%%%%%%%%%%%%%%%%%%%%%%%%%%%%%%%%%%%%%%%%%%%

% Don't change these lines
\bsp	% typesetting comment
\label{lastpage}
\end{document}